\documentclass[preprint,12pt,authoryear]{elsarticle}

\usepackage{amssymb}
\usepackage{amsmath}
\usepackage{makecell}
\usepackage{graphicx}
\usepackage{subcaption}
\usepackage{booktabs}
\usepackage{multirow}
\usepackage{threeparttable}
\usepackage[hyphens]{url}
\usepackage[hidelinks]{hyperref} 
\usepackage{adjustbox}
\usepackage{siunitx}
\journal{Computers in Human Behavior}

\begin{document}

\begin{frontmatter}




\title{The Decision to Verify: How Warmth and User Characteristics Shape Reliance on Conversational Agents for Information Search}

\affiliation[label1]{organization={Amsterdam University of Applied Sciences},
             addressline={Fraijlemaborg 133},
             city={Amsterdam},
             postcode={1102 CV},
             country={Netherlands}}
\affiliation[label2]{organization={Leiden University},
             addressline={Einsteinweg 55},
             city={Leiden},
             postcode={2333 CC},
             country={Netherlands}}
             
\author[label1,label2]{Mert Yazan}
\ead{m.yazan@hva.nl}

\author[label1]{Frederik Bungaran Ishak Situmeang}
\ead{f.b.i.situmeang.hva.nl}

\author[label2]{Suzan Verberne}
\ead{s.verberne@liacs.leidenuniv.nl}

\begin{abstract}


Conversational artificial intelligence (AI) provides an efficient and convenient gateway to information access. However, it can cause overreliance when users blindly trust AI and accept its answers without fact-checking. Information search increasingly follows a hybrid interaction paradigm that combines conversational AI with web search, making fact-checking easier. In this paper, we examine whether this interaction paradigm is effective in curbing reliance. We further investigate the underlying factors (e.g., digital literacy and conversation warmth) that drive users to verify AI answers. We conduct a mixed-subjects question-answering experiment where participants interact with either a warm or a neutral chatbot. Our findings reveal that reliance persists despite users having access to both conversational and web search. The decision to verify is driven primarily by existing user perceptions (e.g., prior trust in chatbots) rather than answer properties, with some users fact-checking regardless of the context and others trusting chatbots by default. Warm conversational style has an indirect yet critical influence on reliance by increasing agreement with the chatbot when it is incorrect. Consulting additional AI sources predicts higher accuracy, while traditional web search does not. Our study extends overreliance research by: (a) demonstrating its persistence despite access to fact-checking, (b) identifying verification behavior as user-dependent, and (c) revealing conversational warmth's indirect effect on overreliance with implications for designing trustworthy conversational search systems.

\end{abstract}

\begin{graphicalabstract}
    \includegraphics[width=1\textwidth]{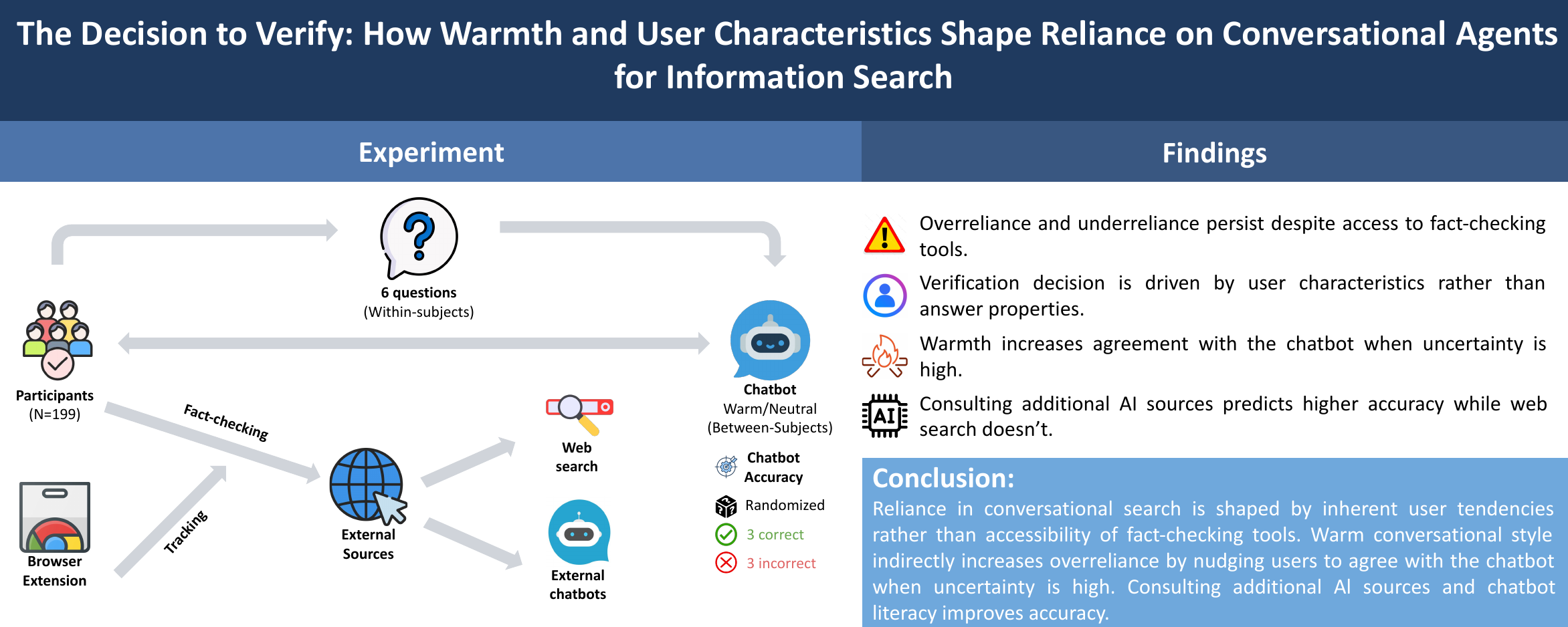}
\end{graphicalabstract}

\begin{highlights}

\item Reliance persists in conversational search despite access to fact-checking tools.
\item The decision to verify chatbot answers is driven by user characteristics.
\item Some users always fact-check; others never do.
\item Warmth leads to reliance by increasing agreement with incorrect chatbot answers.
\item Consulting additional AI sources predicts higher accuracy.

\end{highlights}

\begin{keyword}
Reliance \sep Large Language Models \sep Chatbots \sep Trust \sep Fact-checking \sep Warmth \sep Information-seeking



\end{keyword}

\end{frontmatter}


\section{Introduction}

    Information search has seen a rapid change in recent years, becoming more conversational as the popularity of Large Language Models (LLMs) has grown \citep{chatterji2025chatgpt}. Approximately 10\% of the world's adult population uses ChatGPT \citep{chatgpt} at least weekly, with one in four queries involving information seeking \citep{chatterji2025chatgpt}. Compared to traditional search interfaces, such as search engines, conversational agents offer better accessibility, are easier to use, and require less effort \citep{user_interact_genir, overreliance_llm_search, new_era_search}.  
    
    Despite the advantages, there exists a major downside of conversational agents: hallucinations. Humans rely on AI outputs even when they are incorrect, showing overreliance \citep{overreliance_risk}. Overreliance leads people to accept the first answer they receive from an LLM without further inquiries in challenging tasks \citep{overreliance_llm_search, Kim2024Certainty}. Furthermore, chatbots give an inflated sense of trust and confidence, even when they perform worse than search engines \citep{ux_chatgpt_google, blending_queries}. LLMs hallucinate responses that seem plausible at first glance, albeit being nonfactual \citep{Huang2025hallucination}, making overreliance a critical issue for conversational agents \citep{ux_chatgpt_google, overreliance_llm_search}. 

    In controlled experiments, participants are generally constrained to interact with either a chatbot or with a search engine \citep{overreliance_llm_search, user_interact_genir}. However, in reality, users can access both: Recent developments have introduced a new search paradigm that provides users with simultaneous access to conversational agents (e.g., chatbots) and web search \citep{google_genai_search}. Due to this new paradigm, hallucinations are not regarded as a major issue for most users, since chatbot responses can be easily verified by external sources \citep{exploring_motivations, chatting_with_chatgpt}. 

    It can be argued that the new paradigm solves overreliance by combining the strengths of both conversational and traditional approaches. However, there is evidence showing that the mere presence of AI can induce reliance \citep{Kim2024Certainty}. Participants seek external confirmation less frequently when interacting with AI and tend to agree more with AI responses regardless of accuracy \citep{Kim2024Certainty}. Moreover, chatbots still hallucinate and give incorrect answers, even with web search \citep{newsguard2025ai}. These findings suggest that the new search paradigm might not be the definitive answer against overreliance. 
    
    Despite verification tools being readily available with the new paradigm, the decision to fact-check or trust an AI answer is up to the user. Almost 90\% of the utterances contain check-worthy claims in information-seeking tasks \citep{joko2025wildclaims}, making it unrealistic for users to verify all responses. Users have to make a conscious decision about which answers need further verification. However, users fail to fact-check results that confirm existing beliefs \citep{Elsweiler2025query} or become anchored to the first information source they encounter \citep{Aslett2024Misinformation}. Consequently, tool availability alone cannot ensure a solution to reliance, as the ultimate barrier remains the human decision-making process itself.

    Previous experience, AI expertise, and digital literacy shape fact-checking behavior by reflecting users' awareness of LLM limitations \citep{exploring_motivations, bias_perception}. Similarly, trust in AI systems determines whether users accept AI outputs as is \citep{overreliance_risk, conv_xai}. Beyond user background, chatbot conversational style, especially its warmth, influences reliance by shaping risk perceptions and emotional trust \citep{rethink_conv_styles, trust_chatgpt_google, financial_advisors}.

    In summary, overreliance appears to persist despite the emerging trend of integrating conversational AI with traditional search. Existing studies cannot conclusively explain why and how overreliance manifests when users have simultaneous access to both modalities \citep{overreliance_llm_search, user_interact_genir}. Addressing this gap requires a deeper understanding of the psychological and contextual factors that shape users’ verification decisions.
    
    In this paper, we investigate reliance on conversational agents for the new paradigm of information search with a mixed-subjects study (n=199). We designed an experiment based on question-answering where participants were given access to a helper chatbot. The chatbot's answer correctness was randomized, with each participant seeing correct answers half of the time. Participants were randomly assigned to a warm or neutral chatbot and were free to use other sources beyond the chatbot, simulating a realistic search experience. Previous work has treated external source usage as a self-reported binary variable (used, not used), leading to noisy results \citep{Kim2024Certainty}. We track participant interactions with a custom browser extension that records visited URLs, enabling us to measure both verification frequency (number of URLs) and verification type (consulting another AI source or web search).

    We look into not only overreliance but also underreliance by examining participant agreement (with the chatbot) and accuracy, given the chatbot's correctness. Furthermore, we conduct behavioral analysis to understand how reliance manifests across different user groups. With this study, we answer the following research questions:

    \begin{itemize}
        \item \textbf{RQ1:} How do overreliance and underreliance manifest in conversational search when users have access to fact-checking tools?
        \item \textbf{RQ2:} What factors drive users’ decisions to verify or accept a chatbot’s answer? 
        \item \textbf{RQ3:} What is the influence of conversation warmth on reliance?
    \end{itemize}
    
    Our findings confirm the existence of both types of reliance despite access to fact-checking tools. Some participants exhibited underreliance by always fact-checking, incorrectly assuming that the questions were too difficult for the chatbot. On the other side, other participants showed overreliance and never felt it necessary to question the chatbot responses. Overreliance also manifested subconsciously through conversational warmth: participants agreed more with the warm chatbot than the neutral chatbot when incorrect answers were provided, likely due to increased uncertainty. Behavioral analysis showed that source preferences also vary across participants, with some consulting additional AI sources and others relying on web search. Finally, chatbot literacy was effective in increasing accuracy when the chatbot was incorrect.

    Our study fills an important gap in the literature by confirming the existence of both overreliance and underreliance in the new search paradigm. Beyond identifying reliance, we further show how it manifests differently across users based on their fact-checking tendencies. Our findings reveal that conversational warmth subconsciously influences reliance behavior by manipulating user agreement with the chatbot's answer. This study provides the first empirical evidence of this indirect yet consequential effect.

\section{Background}

    \subsection{Conversational Search}
    \label{adoption}

    
    In terms of performance, conversational search does not significantly outperform traditional web search \citep{user_interact_genir, new_era_search}. However, it reduces user effort, as LLM-based search is rated as more satisfactory and requires fewer queries to complete tasks \citep{overreliance_llm_search}. \citet{new_era_search} asked users to perform a search task and select the optimal option among several alternatives, either by using a search engine (Google) or a chatbot (ChatGPT), while allowing users to visit external websites. Chatbot users visited fewer external websites, without this negatively impacting their search performance \citep{new_era_search}. 
    
    When users interacted with both Bing Chat and Bing Search across 24 tasks, their satisfaction was significantly higher with Bing Chat across all tasks \citep{user_interact_genir}. Objectively, performance did not differ, but users felt more successful with Bing Chat and exercised less effort (e.g., number of clicks, duration) \citep{user_interact_genir}. Participants who received assistance from a chatbot for fact-checking verified claims faster than those who read Wikipedia passages, while achieving similar accuracy \citep{Si2024Large}.

    \subsection{Trust and Overreliance}
    \label{trust_and_overreliance}

    The practical benefits of conversational agents increase user trust and ease the adoption process \citep{impact_of_trust, exploring_intention, exploring_trust}. On the other hand, trust is also a predictor of reliance, and consequently, overreliance \citep{overreliance_risk, conv_xai}. Overreliance is defined as accepting incorrect AI predictions \citep{Kim2024Certainty, hashemi2024evaluating, overreliance_llm_search}. For instance, users may exhibit overreliance by following AI advice over expert recommendations even when it contradicts clear evidence, despite negative consequences for themselves and others \citep{overreliance_risk}. Conversely, underreliance occurs when users reject correct AI predictions \citep{hashemi2024evaluating}.

    LLMs are prone to hallucinate plausible yet nonfactual content because of their generative nature \citep{Huang2025hallucination}, exacerbating the risk of overreliance. While the average task accuracy in controlled experiments yields similar results between LLMs and web search \citep{user_interact_genir, new_era_search}, this does not reflect the reliance risk associated with AI: Users who received explanations from ChatGPT had lower accuracy for incorrect answers \citep{Si2024Large}, and when answering medical questions, people with AI access fact-checked significantly less and underperformed \citep{Kim2024Certainty}.
        
    In the context of information-seeking, people who interacted with ChatGPT reported higher trust despite inconsistent answers, compared to those who interacted with Google \citep{ux_chatgpt_google}. When conducting a series of consumer product research tasks using either an LLM-based search tool or a traditional search engine, the LLM group tended to accept the first answer they received, even in challenging tasks with low accuracy \citep{overreliance_llm_search}. More worryingly, people who primarily interacted with chatbots showed higher confidence even when their accuracy was low: Several participants described the chatbot interface as more reliable than web search because its answers ``sounded reasonable'' or ``made sense'' \citep{blending_queries}. 

    Despite being understudied, underreliance can be as harmful as overreliance. Some users show hesitation to accept model predictions, even when they are substantially more accurate than humans \citep{hashemi2024evaluating}. Despite models providing explanations, underreliance persisted, showing an active avoidance or skepticism towards AI \citep{hashemi2024evaluating}.

    \subsection{Fact-checking Behavior}

        \subsubsection{A New Search Paradigm}
        
        Section~\ref{trust_and_overreliance} establishes the existence of reliance: Users interacting exclusively with conversational agents are more prone to overreliance than those using search engines \citep{ux_chatgpt_google, overreliance_llm_search, user_interact_genir}. However, in reality, users can access conversational and web search interfaces simultaneously, reflecting a new paradigm of search \citep{google_genai_search}. In fact, users rarely raise concerns about factual correctness or hallucinations due to the utility of conversational agents \citep{exploring_motivations}, and consider hallucinations to be mitigable since chatbot responses can be fact-checked \citep{chatting_with_chatgpt}. 
        
        On top of simultaneous access, a more unified searching experience has emerged: Most search engines include AI-generated summaries, even placing them above the search results \citep{google_genai_search}, and chatbots search the web before generating an answer \citep{openai2024chatgptsearch}. One might argue that this unification of functionality practically eliminates reliance, since chatbots can fact-check themselves and provide users with easy access to fact-checking tools.
    
        In practice, the new, unified search paradigm does not solve hallucinations: chatbots continue to provide incorrect answers even after web search \citep{newsguard2025ai}. On the user side, access to fact-checking tools does not necessarily eliminate overreliance. \citet{Kim2024Certainty} examined how AI uncertainty expressions affected user behavior in a medical question-answering task in which participants were free to consult external sources besides the experiment chatbot. Web search reduced blind agreement and led to accuracy gains; however, overall accuracy was still lower than in the no-AI condition.

        \subsubsection{Cognitive Biases}
    
        Regardless of the paradigm or access to external tools, the decision to fact-check is up to the user. Yet, this decision is clouded by cognitive biases: users tend to favor results that confirm their beliefs \citep{Elsweiler2025query}, meaning an answer that reinforces existing beliefs might not be further investigated. Conversational presentation makes information seem more credible than static text (e.g., Wikipedia) and hurts the ability to detect inaccuracies \citep{Anderl2024Conversational}. Fact-checking is also not exempt from bias: across five experiments on false news identification, researchers found that searching increased belief in misinformation and raised the probability of rating false/misleading articles as true by 18–22\% \citep{Aslett2024Misinformation}. 
    
        Given cognitive biases, it is unrealistic to expect users to correctly assess which answers need fact-checking, especially when 88\% of utterances contain check-worthy claims in information-seeking tasks \citep{joko2025wildclaims}. While \citet{Kim2024Certainty}'s findings suggest that overreliance persists despite access to external sources, their study had several limitations: it focused on uncertainty expression rather than reliance, did not investigate why or how users interacted with external sources, relied on noisy self-reported measures of external search, and excluded underreliance. We take their study one step further with a more precise participant tracking system and claim the following hypothesis:
        
        \textbf{H1:} Being able to fact-check does not eliminate overreliance or underreliance in conversational search.

    \subsection{Individual Background on Fact-checking}

    Besides the cognitive biases, reliance on AI systems may be rooted in individual characteristics that users bring to their interactions: According to \citet{exploring_motivations}, one reason users overlook the limitations of conversational interfaces might be their lack of expertise and digital literacy. People with expertise in AI are more aware of the limitations of natural language generation \citep{bias_perception}. Users with higher baseline trust in automation were predisposed to accept AI outputs with less scrutiny, while those with lower trust dismissed correct information unnecessarily \citep{overreliance_risk, conv_xai}. Regarding verification with web search, individuals with lower digital literacy were especially prone to encountering unreliable results \citep{Aslett2024Misinformation}.

    Prior research has identified overreliance but not examined its underlying causes \citep{Kim2024Certainty, overreliance_risk}. Understanding whether fact-checking decisions stem from personal factors, such as prior knowledge, technology literacy, or trust, is crucial for discovering the true causes of reliance. Since users favor answers that confirm their existing beliefs \citep{Elsweiler2025query}, we form the following hypothesis:
                
    \textbf{H2:} Prior attitudes toward conversational AI systematically shape verification behavior.

    \subsection{Chatbot Conversation Style}

    Chatbots often employ a friendly, positive personality by exhibiting high openness, conscientiousness, and extraversion \citep{PsychoBench}, and are designed to appear human-like through anthropomorphic features such as avatars and emotional responses \citep{anthro_cues}. According to the Stereotype Content Model \citep{SCM}, a social psychology theory, stereotypes form along two dimensions: warmth and competence. While competent-sounding chatbots increase technology acceptance \citep{rethink_conv_styles}, a predictor of trust \citep{chatgpt_we_trust, exploring_intention, exploring_trust}, the influence of warmth is more complex and context-dependent.
    
    Research on warmth in chatbot interactions has produced mixed findings. \citet{rethink_conv_styles} found that warmth did not increase acceptance of chatbots. However, other findings suggest warmth-related features may influence behavior through indirect mechanisms: high anthropomorphism increased forgiveness of chatbot errors \citep{chatbot_forgiveness}, perceived human-likeness predicted trust in ChatGPT \citep{trust_chatgpt_google}, and risk perception decreased with warmer chatbots \citep{rethink_conv_styles}. Most notably, \citet{financial_advisors} demonstrated that users trust an extroverted LLM financial advisor more, despite objectively worse guidance. We propose that warmth influences reliance subconsciously and formulate the following hypothesis:

    \textbf{H3:} A warm conversation style increases user agreement with chatbot answers, thereby amplifying overreliance.

\section{Method}

To answer our research questions and evaluate our hypotheses, we conduct a mixed-design experiment in which participants answer six multiple-choice questions (within-subjects) using either a warm or neutral chatbot (between-subjects). The survey was conducted via Qualtrics, and participants were recruited from Prolific. The study was ethically approved by the University of Amsterdam Economics and Business Ethics Committee with approval number EB-18789. In the following subsections, we provide a detailed explanation of our experimental design.

    \subsection{Experiment Design}

        \subsubsection{Question Answering Task}

        Participants answered six multiple-choice questions presented in randomized order. Each question page included an embedded chatbot for assistance. While participants could choose whether to interact with the chatbot, we displayed its answer by default to ensure exposure. To confirm participants had seen the chatbot's answer, we activated the next page button after ten seconds and included an attention check asking which option the chatbot selected. Before beginning the experiment, participants completed a tutorial introducing them to the flow. As an incentive, participants received £0.10 per question answered.

        We designed two chatbots with different conversational styles: neutral and warm. Participants were randomly assigned to one chatbot for the entire experiment. We employed a counterbalanced within-subjects design for answer correctness. Each participant encountered both correct (n=3) and incorrect (n=3) responses across six questions, with randomized assignment determining which questions appeared in each correctness condition. This randomization was constrained to maintain roughly equal distribution of correctness conditions across all six questions at the sample level. This correctness randomization approach follows methodology from medical decision-making research \citep{gaube2021do}.
    
        \subsubsection{Questions}

        \begin{table}[t]
            \scriptsize
            \begin{threeparttable}
            \begin{tabular}{|l|l|l|l|}
            \hline
            \textbf{Index} & \textbf{Question} & \textbf{Correct Answer} & \textbf{Incorrect Answer} \\
            \hline
            Q1 & \makecell[l]{Which global investment firm recently \\ committed a \$5 billion India-focused \\ fund in 2025?} & None & BlackRock \\
            \hline
            Q2 & \makecell[l]{Which country has the world’s largest \\ forested area per capita?} & 
            Suriname\tnote{1} & French Guiana \\
            \hline
            Q3 & \makecell[l]{In what year did a country first recognize \\ internet access as a basic human right?} & 2009 & 2000 \\
            \hline
            Q4 & \makecell[l]{Which non-primate animals were the first \\ to show mirror self-recognition?} & Bottlenose dolphins & Asian elephants \\
            \hline
            Q5 & \makecell[l]{Which South American country has the \\ second-highest plant biodiversity after \\ Brazil?} & Colombia & Venezuela \\
            \hline
            Q6 & \makecell[l]{Which single man-eating animal is \\ responsible for the most deaths ever?} & Champawat tigress & Anopheles mosquito \\
            \hline
            \end{tabular}
        \begin{tablenotes}
        \scriptsize
        \item[1] This answer was valid during our experimentation period. However, as of November 29th, Guyana has the world's largest forested area per capita. Source: \url{https://www.visualcapitalist.com/mapped-countries-with-the-most-forest-area-per-capita/}
        \end{tablenotes}
        
        \caption{List of questions and the correct and incorrect answers chatbots presented.}
        \label{tab:questions}
        \end{threeparttable}
        \end{table}

        Information retrieval literature identifies different types of information-seeking behaviors based on users' goals and task structures \citep{broder_taxonomy, Belkin1982AskPartI}. For this experiment, we focused on informational questions: those seeking specific facts with verifiable answers.
    
        The first question was taken from Search Arena \citep{searcharena2025} (Q1, see Table~\ref{tab:questions}). Search Arena is part of the LMArena, an open platform created by researchers to compare LLMs based on human preference judgments, asking users to choose between the answers of two search-enhanced LLMs. Q1 poses \textit{Which global investment firm recently committed a \$5 billion India-focused fund in 2025?}, while no actual firm has made such a commitment.
            
        We initially planned to create the question set entirely from questions where both LLM answers were labeled as bad at Search Arena, like Q1. However, we found that current LLMs could correctly answer most questions previously labeled as bad. While LLMs may have struggled with these questions when the dataset was created, iterative improvements have since enabled them to answer correctly. Recognizing that similar datasets face the same problem, we decided to craft the remaining questions ourselves based on our analysis of what made Q1 challenging.

        LLMs struggled with Q1 primarily because multiple sources provided partially correct answers that failed to satisfy all criteria simultaneously (e.g., investments made in 2025 but not by an investment firm, or investments by investment firms but not in 2025). Verifying whether each source aligns with question conditions proved challenging for LLMs, leading to inconsistent responses. Our investigation revealed that chatbots occasionally provided correct answers, but more often gave incorrect ones.
        
        Given what we identified, we created the remaining five questions. For each question, we checked search engine results and various sources to understand what makes it difficult. We noticed that, similar to Q1, chatbots fail to answer our questions due to a lack of attention to detail and confusion due to multiple conflicting sources. The full list of questions and the correct and incorrect answers are shown in Table~\ref{tab:questions}. We further analyze the common pitfalls chatbots and search engines fall into for each question in~\ref{question_pitfalls}. The final question set covered diverse domains, i.e., finance, geography, law, and biology, with some domains represented by multiple questions.
        
        We validated the questions through a pilot study, which revealed natural variation in difficulty. This variation was intentional: our goal is to observe reliance behavior across different question contexts. A participant who exhibits consistent reliance patterns, regardless of question difficulty, provides valuable information about their underlying behavioral tendencies. Additionally, we used multi-level models with questions nested as a random effect, which statistically accounts for any systematic differences between items.

        Three of our questions concerned past events, whereas the remaining questions were subject to change over time. Throughout the data collection period, we monitored whether any answers had changed. Although no changes occurred during participant recruitment, we observed that the sources underlying one question (Q2: Which country has the world's largest forested area per capita?) were updated after the experiment had concluded. This does not affect our results, as all data collection had already been completed.

        \subsubsection{Answer Pre-generation}
        \label{answer_generation}

        \begin{figure}[!ht]
            \centering
            \begin{subfigure}{0.7\textwidth}
                \centering
                \includegraphics[width=\linewidth]{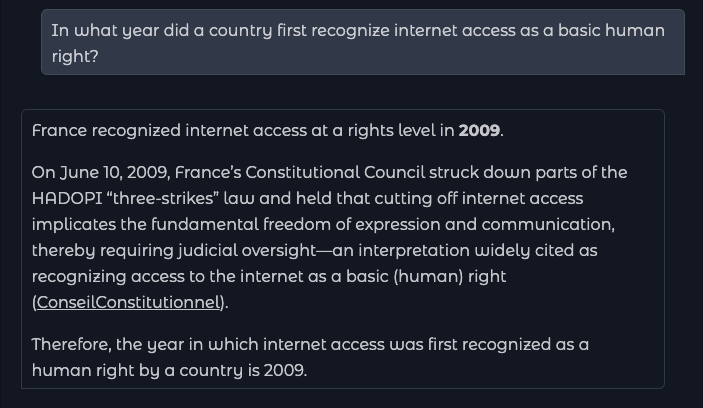}
                \caption{Neutral answer, without any emojis or emotional expressions.}
                \centering
                \includegraphics[width=\linewidth]{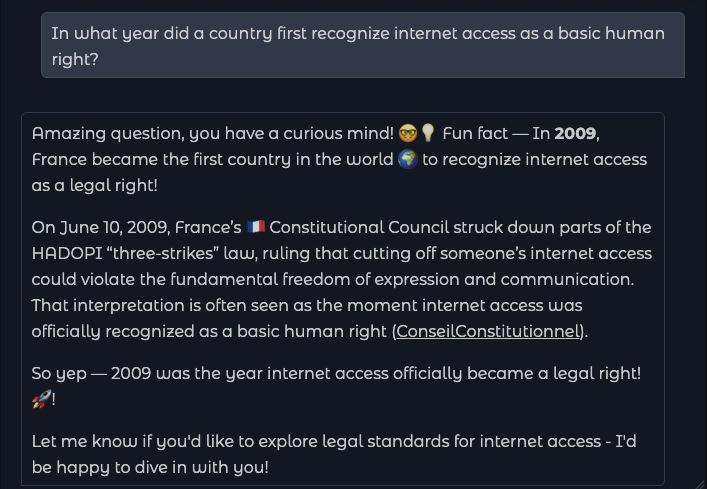}
                \caption{Warm answer: friendly tone with emojis and emotional expressions.}
            \end{subfigure}
            \caption{Comparison of correct neutral and warm answers for Q4.}
            \label{fig:chat_answers}
        \end{figure}

        As mentioned in previous sections, for each question, we initiated the conversation with the chatbots by pre-generating their answers. This was implemented to ensure participants see the chatbot's answer, even when they do not interact with it. In total, four answers were generated for each question based on correctness and warmth. First, we generated neutral answers by sampling the neutral chatbot (temperature = 0.7) to obtain a correct and an incorrect answer. To make our study realistic, we let the experiment chatbots conduct web search. We noticed that sometimes LLMs provide non-working links. We continued sampling until we reached responses with only working links.        

        To ensure conversational style was the only difference between conditions, we did not generate warm answers independently. Instead, the warm chatbot rewrote the neutral answers in its warm style. Then we applied post-processing so that all warm answers followed the same structure: a warm introduction that says ``Amazing question!'', followed by a warm answer with emotional words and emojis. Finally, a friendly closure that says: ``Let me know if you'd like to explore [TOPIC] further - I'd be happy to dive in with you!''.  See Figure~\ref{fig:chat_answers} for an example showing the difference between a neutral and a warm answer.

        We have observed that the answer lengths did not vary significantly between questions, and we believe it is beneficial to have small variations between questions, as we will be treating them independently. Yet, we did not want the structure of correct and incorrect answers to be too different, as it might introduce confounders. Therefore, we modified the answers by removing a couple of words or links to make the correct and incorrect answers of a question similar in structure. 

        Each question was presented with four multiple-choice options. One option represented the accurate answer, the one provided when the chatbot is correct. The incorrect answer the chatbot provided was fixed; it was the same answer for everyone, among the three incorrect choices. We manually selected the most common misleading answer from search results as the fixed incorrect response provided by the chatbot. The remaining two multiple-choice options were also manually drawn from common, misleading search results to serve as plausible distractors.
        

        \subsubsection{Fact-checking}
       
        Our experiment is designed to replicate how people search for information in real life, where they have access to multiple tools. That is why, in parallel to providing participants with an embedded chatbot, we also placed a ``Web Search'' button in the question pages. This button automatically opens a new tab with Google Search, which shows the search results for that question and encourages participants to fact-check. 

        We nudged participants to stay in the survey platform and use the tools we provided. However, we did not enforce them: Given our experimental design, we wanted participants to approach the search naturally, with any tool of their choice. Therefore, we permit them to use other sources (other chatbots or search engines).

            \paragraph{Participant Tracking}

            To track the sources participants interacted with, the simplest option was to ask them after each question whether they had used an external source \citep{Kim2024Certainty}. Despite its implementation simplicity, this approach was not desirable for two reasons: a) it increased the chance of survey fatigue, and b) it risked not properly tracking which sources participants had interacted with if they decided not to share. Given these risks, we have created a Chrome extension designed to track participants.\footnote{\url{https://chromewebstore.google.com/detail/qualtrics-url-tracker/bkihikddhlacccpjekjgplocinicnlmf?authuser=0&hl=en&pli=1}}

            The extension is limited to tracking which tab the participant is active on. It does not track cursor movements nor the participant interactions within the tab. The extension transparently shows which URLs were tracked and lets participants remove specific URLs. Tracking is only active during the experiment; it starts when the first question is encountered, and it automatically stops after the participant sees the last question. At the beginning of the survey, we asked for consent and informed participants how to install the extension. We emphasized that they were expected to install it. After the experiment was over, we asked participants to share the Qualtrics response ID generated by the extension.

            \paragraph{AI-Generated Search Summaries}
            
            Initially, we used DuckDuckGo Lite for the ``Web Search'' button since it does not provide AI summaries. However, after a pilot, we noticed that participants abandoned the web search button and used Google instead. We therefore decided to switch to Google to reduce participants' labor and directed them to the ``web'' tab of Google Search, which shows the search results without an AI summary. Participants were free to check the AI summaries afterwards, as this was tracked as a new URL, too. We noticed that the layout and the nature of Google AI Mode and AI summaries change frequently. To mitigate the risk of participants encountering different layouts, we monitored the interface for any changes throughout the experiment period.

        \subsubsection{Survey Flow}

        \begin{figure}[!ht]
            \centering
            \includegraphics[width=\linewidth]{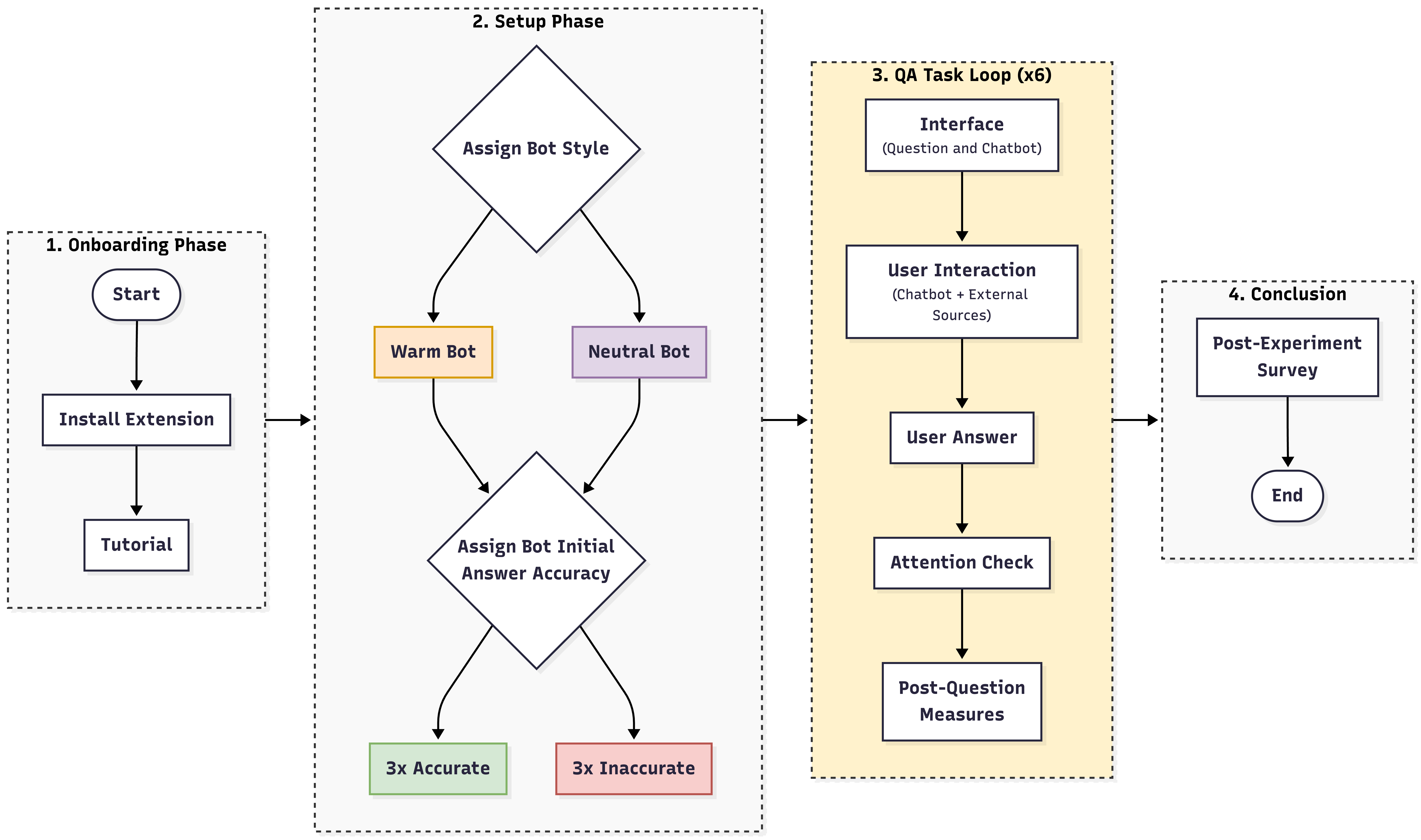}
            \caption{An overview of the survey flow. Chatbot style and answer correctness were randomly assigned once per participant, as well as the question order in the task loop.}
            \label{fig:survey_flow}
        \end{figure}

        Figure~\ref{fig:survey_flow} outlines our survey flow. Participants started the survey with a consent form and an explanation about how to install the browser extension. Next, they were introduced to the question answering task. Before starting with the first question, they had a tutorial session on the chatbot and the ``Web Search'' button. The task began when the participant encountered the first question. After answering each question, participants received an attention check asking them to choose the answer the chatbot suggested. Afterwards, they filled out three items to measure their confidence, familiarity with the topic, and perceived difficulty of the question for the chatbot. This was repeated for all questions.

        After completing the task, participants were asked to provide the response ID generated by the browser extension. Then, they filled a perceived warmth scale to assess the manipulation check, followed by trust and literacy scales. We posed a mandatory open-ended question (\textit{``How did you decide whether to trust or doubt the chatbot’s answers?''}) afterwards to gather qualitative insights. Finally, the survey is concluded with demographic items.
        
        \subsubsection{Technical Setup}
        \label{technical_setup}

        \begin{figure}[!ht]
            \centering
            \begin{subfigure}{0.45\textwidth}
                \centering
                \includegraphics[width=\linewidth]{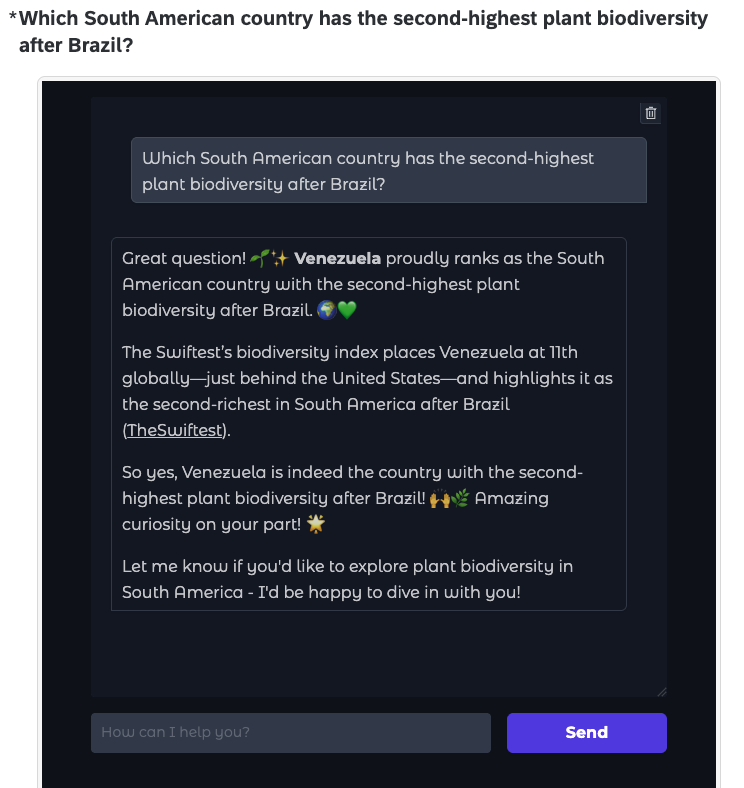}
                \caption{The participant first sees the question and the chatbot's answer.}
            \end{subfigure}
            \hfill
            \begin{subfigure}{0.45\textwidth}
                \centering
                \includegraphics[width=\linewidth]{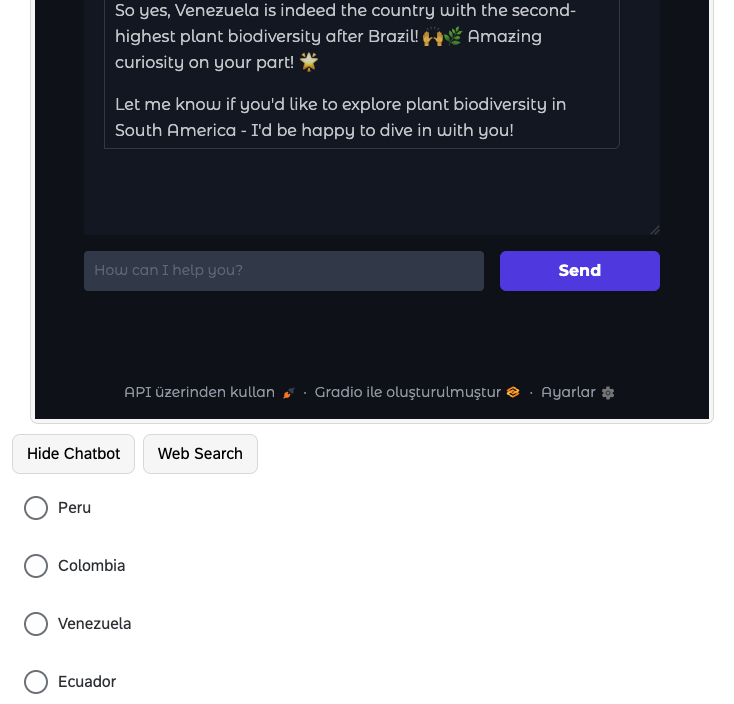}
                \caption{Below the chatbot container, there are 2 buttons (``Hide Chatbot'' and ``Web Search'').}
            \end{subfigure}
        
            \vspace{0.5cm} 
        
            \begin{subfigure}{0.5\textwidth}
                \centering
                \includegraphics[width=\linewidth]{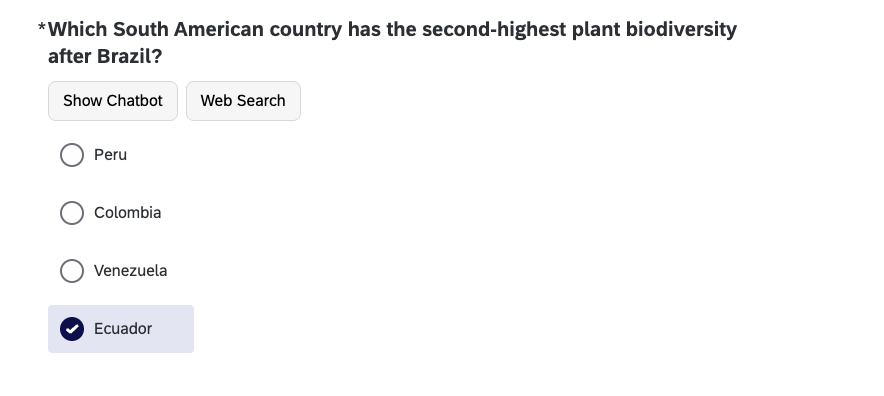}
                \caption{``Hide Chatbot'' hides the chatbot container. ``Show Chatbot'' makes it visible again.}
            \end{subfigure}
            \hfill
            \begin{subfigure}{0.45\textwidth}
                \centering
                \includegraphics[width=\linewidth]{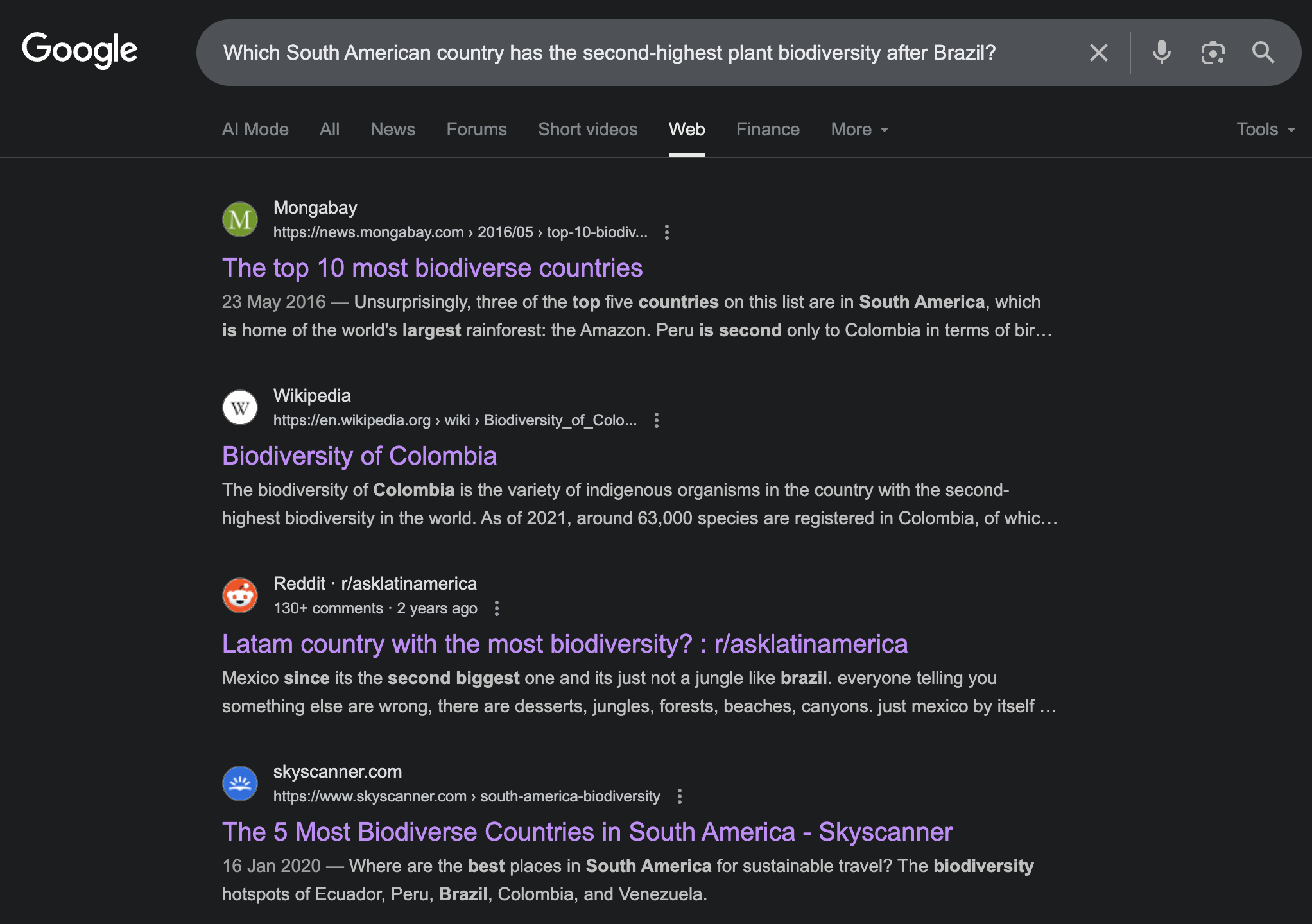}
                \caption{``Web Search'' directs participants to the search results.}
            \end{subfigure}
        
            \caption{The platform we created for the experiment, embedded in question pages.}
            \label{fig:platform}
        \end{figure}
        
        To design the chatbot platform, we used Gradio \citep{abid2019gradio}. Our chatbot was based on GPT-4.1, accessed through the OpenAI API. While generating initial answers, we set the temperature to 0.7 to obtain both correct and incorrect answers. The temperature is then set to 0 for future interactions. The API allows developers to choose which tools the model can use. The web search tool was enabled in the API both for initial answers and subsequent interactions.

        We instructed the LLM via the system prompt to generate short answers and always provide links to the sources it has used. For the warm chatbot, we explicitly stated that ``You are a warm, friendly assistant.'' and ``You use emojis and human-like language.'' to increase its warmth and human-likeness. We noticed that GPT 4.1's default personality is fairly neutral when answering factual questions; therefore, we did not explicitly instruct it on conversation style.
                
        The browser extension was created using JavaScript and the Flask framework\footnote{\url{https://flask.palletsprojects.com/en/stable/}}. The chatbot platform was embedded inside the question pages in Qualtrics using JavaScript and HTML/CSS. Participants could toggle the chatbot platform's visibility using a button placed side by side with the web search button. Since we wanted participants to see the chatbot answers, we made it visible by default. Whether the participant clicked on the ``Hide/Show Chatbot'' or ``Web Search'' buttons was tracked, using custom JavaScript code embedded in Qualtrics. Figure~\ref{fig:platform} shows the platform we created. 
        
    \subsection{Measuring Reliance}
    \label{measuring_reliance}

    Previous studies have used various variables (i.e., accuracy, confidence, trust) as proxies to identify overreliance \citep{overreliance_llm_search, plan_then_execute, overreliance_risk}. \citet{Kim2024Certainty} used a combination of dependent variables and behavioral measures. Since their experimental design is the most similar to ours, we investigate reliance in the following ways, influenced by them:

    \begin{enumerate}
        \item How participant accuracy varies depending on the chatbot's correctness.
        \item Whether participants choose the same answer as the chatbot, regardless of the chatbot's correctness.
        \item Investigating fact-checking behavior in conjunction with the chatbot's correctness.
    \end{enumerate}

    \subsection{Question-Level (Within-Participant) Variables}

    Question-level, within-participant variables represent measures asked after each question (e.g., participant's confidence or familiarity with the question), behavioral measures (e.g., how many URLs the participant has visited before answering the question), or dependent variables (accuracy, agreement).
    
        \subsubsection{Participant Perceived Variables}
        
        After each question, participants were asked about their confidence in their answer (``Confidence'', between 1 and 10), their familiarity with the topic (``Familiarity'', 7-point Likert agreement scale), and how difficult they believed the question was for the chatbot (``Difficulty\textsubscript{chatbot}'', 7-point Likert agreement scale). 

        \subsubsection{Behavioral Variables}
        
        We collected behavioral data by tracking chatbot interactions and participants' information-seeking behavior. Previous work has used behavioral data as a binary indicator based on self-reports of whether participants consulted additional sources \citep{Kim2024Certainty}. Our setup enabled more granular measurement: we logged all chatbot interactions directly, while a browser extension tracked participants' movements between tabs to identify which URLs they opened. We operationalized both variables as counts: the number of conversational turns with the chatbot and the number of URLs visited.

        We categorized URLs into two types based on their nature: AI-related sources (chatbots or AI-generated summaries) and Non-AI websites (all other URLs). Although it would be possible to further distinguish non-AI URLs by flagging those provided by the chatbot, these represented only 1\% of all Non-AI URLs, so we did not create a separate category. The browser extension automatically tracked new tabs opened via the web search button, making a separate category unnecessary. We formally define 4 behavioral variables:
        \begin{enumerate}
            \item InteractionLen: Number of messages sent to the chatbot by the participant.
            \item AI-URLsNum: Number of URLs visited that represent AI solutions such as chatbots, search engine AI-summaries, or search engine AI-mode.
            \item NonAI-URLsNum: Number of URLs visited that are not AI-URLs.
            \item URLsNum: Total number of URLs visited, sum of AI-URLsNum and NonAI-URLsNum. Only used for descriptive purposes and not included in the analysis.
        \end{enumerate}

        Descriptive statistics for all behavioral variables are presented in Table~\ref{tab:behavioral_desc}.

        \subsubsection{Dependent Variables}
        
        We define two dependent variables: accuracy and agreement. ``Accuracy'' is a binary variable representing whether the participant chose the correct answer. ``Agreement'' is a binary variable indicating whether the participant's final answer matched the chatbot's suggested answer.

    \subsection{Participant-Level (Between-Participant) Variables}

    Between-participant variables are asked for or manipulated only once throughout the experiment. They do not depend on the questions and represent conditions (warmth) and participant-specific characteristics (e.g., digital literacy, trust in automation).

        \subsubsection{Warmth Condition}
        
        ``WarmthCondition'' represents the between-subjects condition of warm and neutral chatbots. We treat it as a binary variable in our analysis, with \textit{0} representing the neutral, baseline chatbot and \textit{1} representing the warm chatbot.

        \subsubsection{Chatbot Literacy}

        There is no specific measure designed to estimate user expertise with chatbots. In this study, we define a new construct called ``Chatbot Literacy'', adapted from Web-Oriented Digital Literacy \citep{web_literacy}. We presented participants with seven AI and chatbot-related items, assessing their familiarity with each on a 5-point scale (1: None, 2: Little, 3: Some, 4: Good, 5: Fully). Chatbot literacy was collected post-experiment. See Table \ref{tab:survey_items} for a list of its items.
            
        \subsubsection{Trust}
    
        Trust is a central construct for studying reliance, specifically as an outcome to measure overreliance \citep{overreliance_risk}. Since we use other dependent variables and behavioral measures (see Section~\ref{measuring_reliance}), we treat trust as a participant-level tendency: trust in chatbots for information search in general, independent of the experiment. We used a 4-item scale adapted from the Trust in Automation scale \citep{jian2000trust} and the Short Trust in Automation scale \citep{mcgrath2025trust}. Responses were measured on a 7-point Likert agreement scale and collected post-experiment. See Table \ref{tab:survey_items} for a list of its items.
        
    \subsection{Control Variables}
    
        To control for experimental design factors, we included three variables. ``Correctness'' is a binary variable about the chatbot's answer correctness. Due to some participants being filtered, the chatbot answer correctness distributions were not equally split (see \ref{question_statistics}). However, they differ marginally (see Figure~\ref{fig:correctness}). Still, we decided to include correctness as a control variable to account for any distribution-imbalance-related effects.
        
        ``SeenOrder'' is a number between 1 and 6 that represents the order in which the question was presented to the participants. Finally, to control the influence of the chatbot's initial correctness, we added ``InitCorrectness'' as a binary variable. Correctness and SeenOrder are within-participant variables, while InitCorrectness is a between-participant variable.
    
    \begin{table}[t]
        \centering
        \scriptsize
        \begin{tabular}{l|l|l|l|l}
        \hline
        \textbf{Variable} & \textbf{\makecell[l]{Participant \\ Level}} & \textbf{Type} & \textbf{Range} & \textbf{Explanation} \\
        \hline
        Confidence & Within & IV & $[1,10]$ & Self-confidence in the answer \\
        \hline
        Familiarity & Within & IV & $[1,7]$ & \makecell[l]{Perceived familiarity with question topic} \\
        \hline
        Difficulty\textsubscript{chatbot} & Within & IV & $[1,7]$ & \makecell[l]{Perceived difficulty of the question for the \\ chatbot to answer} \\
        \hline
        InteractionLen & Within & IV & $\mathbb{N}_0$ & \makecell[l]{The number of messages sent to chatbot} \\
        \hline
        AI-URLsNum & Within & IV & $\mathbb{N}_0$ & \makecell[l]{The number of URLs visited that relate \\ to AI tools (chatbots, AI-summaries)} \\
        \hline
        NonAI-URLsNum & Within & IV & $\mathbb{N}_0$ & \makecell[l]{The number of URLs visited that do not \\ relate to AI tools} \\
        \hline
        Accuracy & Within & DV & $[0,1]$ & \makecell[l]{Accuracy of the participant's answer} \\
        \hline
        Agreement & Within & DV & $[0,1]$ & \makecell[l]{Choosing the same answer as the chatbot }\\
        \hline
        Correctness & Within & CV & $[0,1]$ & \makecell[l]{Correctness of the chatbot's initial answer} \\
        \hline
        SeenOrder & Within & CV & $[1,6]$ & \makecell[l]{The order the question was encountered} \\
        \hline
        InitCorrectness & Between & CV & $[0,1]$ & \makecell[l]{Correctness of the chatbot's answer to the \\ first question participant encountered} \\
        \hline
        WarmthCondition & Between & IV & $[0,1]$ & \makecell[l]{Chatbot warmth condition; neutral or warm} \\
        \hline
        ChatbotLiteracy & Between & IV & $[1,5]$ & \makecell[l]{Participant literacy on AI and chatbots} \\
        \hline
        Trust & Between & IV & $[1,7]$ & \makecell[l]{Participant baseline trust towards chatbots} \\
        \hline
        \end{tabular}
        \caption{Overview of all the variables used in the analysis. CV represents control variables.}
        \label{tab:variables_overview}
    \end{table}

    \subsection{Analysis}
    
    We used linear mixed models (LMM) in our analysis to account for participant-level (PID) and question-level (QID) random effects. We fit models as $\mathrm{DV} \sim \mathrm{IVs} + \mathrm{CVs} + (1 \mid \mathrm{PID}) + (1 \mid \mathrm{QID})$ for both of our dependent variables. IVs and CVs represent all the independent and control variables listed in Table~\ref{tab:variables_overview}. For WarmthCondition, the neutral chatbot served as the reference category, meaning the coefficient estimates represent the effect of the warm chatbot relative to the neutral baseline. Continuous variables were standardized. LMMs are implemented using the lme4 \citep{lme4} library in R.

    \subsection{Participants}
        
    Initially, we collected data from 231 participants. For data quality, we applied multiple filtering steps. First, we filtered based on attention checks: We asked participants to choose which answer the chatbot had given after each question. Those who failed more than three attention checks (>50\%) were rejected. Second, we identified participants who did not put in a meaningful effort: If the participant had chosen the same answer as the chatbot for all questions, and if they finished the survey under $<300$ seconds (1.5 standard deviations below the mean, $M_{\text{CompletionTime}} = 1197$ seconds, $SD_{\text{CompletionTime}} = 602$ seconds), they were filtered. Finally, we excluded participants who exhibited straightlining behavior, providing nearly identical responses across all questions for familiarity, confidence, and difficulty (SD < 1), while simultaneously showing no engagement with either the chatbot or external sources.
    
    After filtering, we had 199 suitable participants. 52.2\% of the group were females, and the age ranged from 19 to 78, with a mean of 45.9. 131 participants (66\%) had a Bachelor's Degree or higher (88 BS/BA, 37 MS/MA, 8 PhD). 46 participants interact with chatbots very actively (3+ times a day), and 58 participants use them 1-2 times a day. \ref{participant_demographics} shows the detailed statistics about the participants. Participants were compensated £12.5/hour for their participation (plus £0.1/question as a reward for accuracy). 

    
    \subsection{Manipulation Check and Item Validity}

    A perceived warmth scale adapted from \citet{rethink_conv_styles} was implemented to check the warmth manipulation (See Table \ref{tab:survey_items} for a list of the items). Warmth was significantly higher ($t(199) = -5.41$, $p < .0001$) for participants in the warm chatbot condition ($M = 5.66$, $SD = 1.04$) than in the neutral chatbot condition ($M = 4.78$, $SD = 1.24$). Despite the difference, we notice that the mean is above average for the neutral condition, showing that people perceive the neutral state of chatbots as warm. Reliability analyses indicated high internal consistency for all multi-item scales (see Table ~\ref{tab:survey_items}): perceived warmth (Cronbach's $\alpha = .94$), trust ($\alpha = .93$), and chatbot literacy ($\alpha = .85$).

\section{Results}

In this section, we first check the correlation between our variables. Then, we share our findings regarding the dependent variables: accuracy and agreement. We further investigate interaction patterns with behavioral variables to understand how reliance manifests. Finally, we conduct a qualitative analysis on the open-ended answers.

    \subsection{Correlation Matrix}

    \begin{figure}[t]
        \centering
        \includegraphics[width=\linewidth]{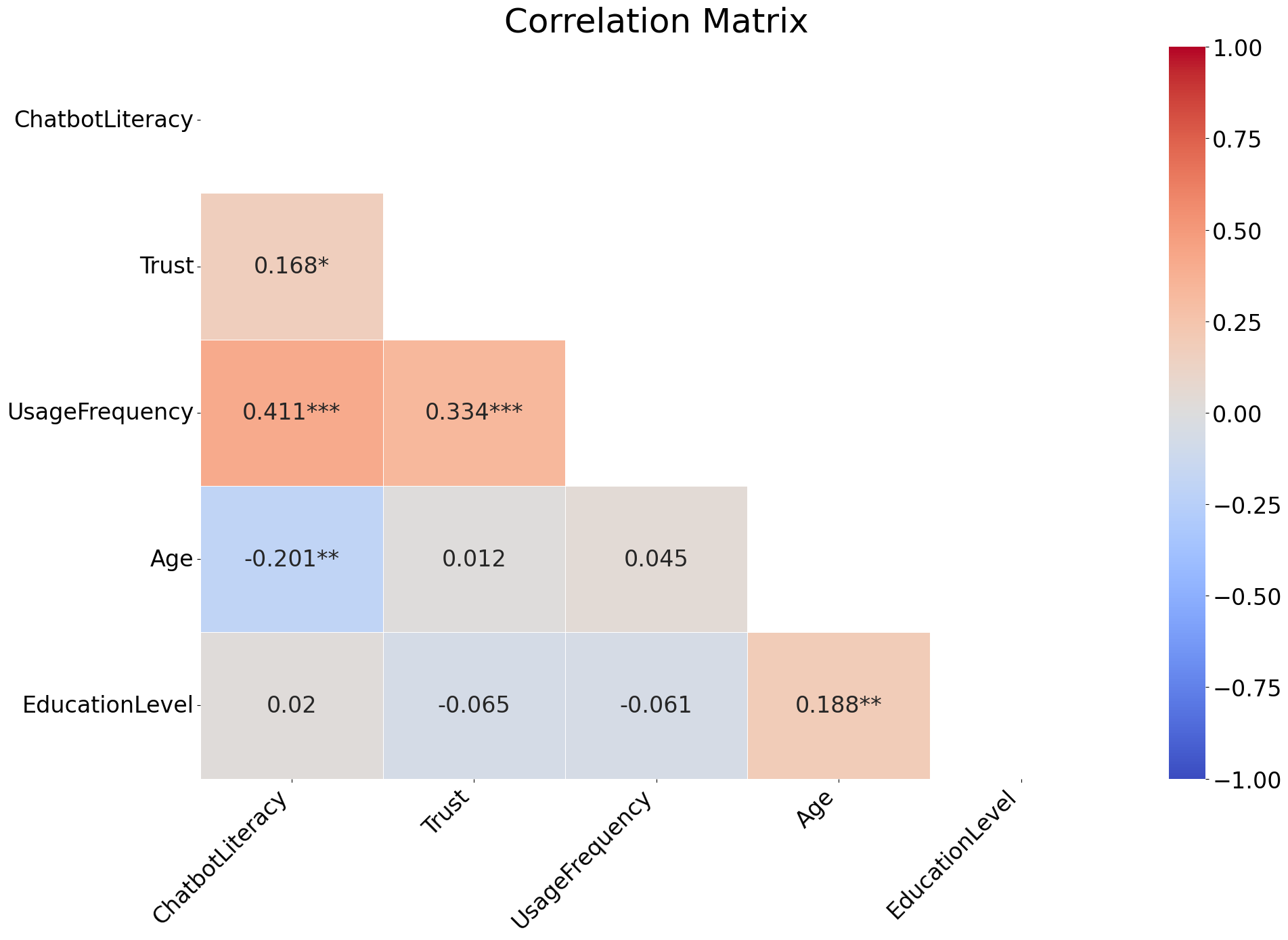}
        \caption{Correlation matrix for post-experiment survey items and demographic variables.}
        \begin{tablenotes}
        \scriptsize
        \item $^{***}p<0.001$, $^{**}p<0.01$, $^{*}p<0.05$
        \end{tablenotes}
        \label{fig:correlation_matrix}
    \end{figure}
    
    Figure~\ref{fig:correlation_matrix} shows the correlation between post-experiment survey measures and demographic variables. People who use chatbots regularly have higher literacy levels ($r=0.411, p < 0.001$) and higher trust in chatbots ($r=0.334, p < 0.001$). To avoid introducing collinearity and to control the number of variables in our model, we thus excluded Usage Frequency from modelling and used it solely for descriptive purposes. Chatbot Literacy is negatively correlated with Age ($r=-0.201, p < 0.01$), and Age is positively correlated with Education Level ($r=0.188, p < 0.01$). Age group has not found to be a significant differentiator of interaction with chatbots for information search \citep{chatterji2025chatgpt}. Therefore, Age and Education Level were excluded for the same reasons, too. 

    \subsection{Accuracy and Agreement}

    \begin{figure}[htbp]
        \centering
        \includegraphics[width=0.7\linewidth]{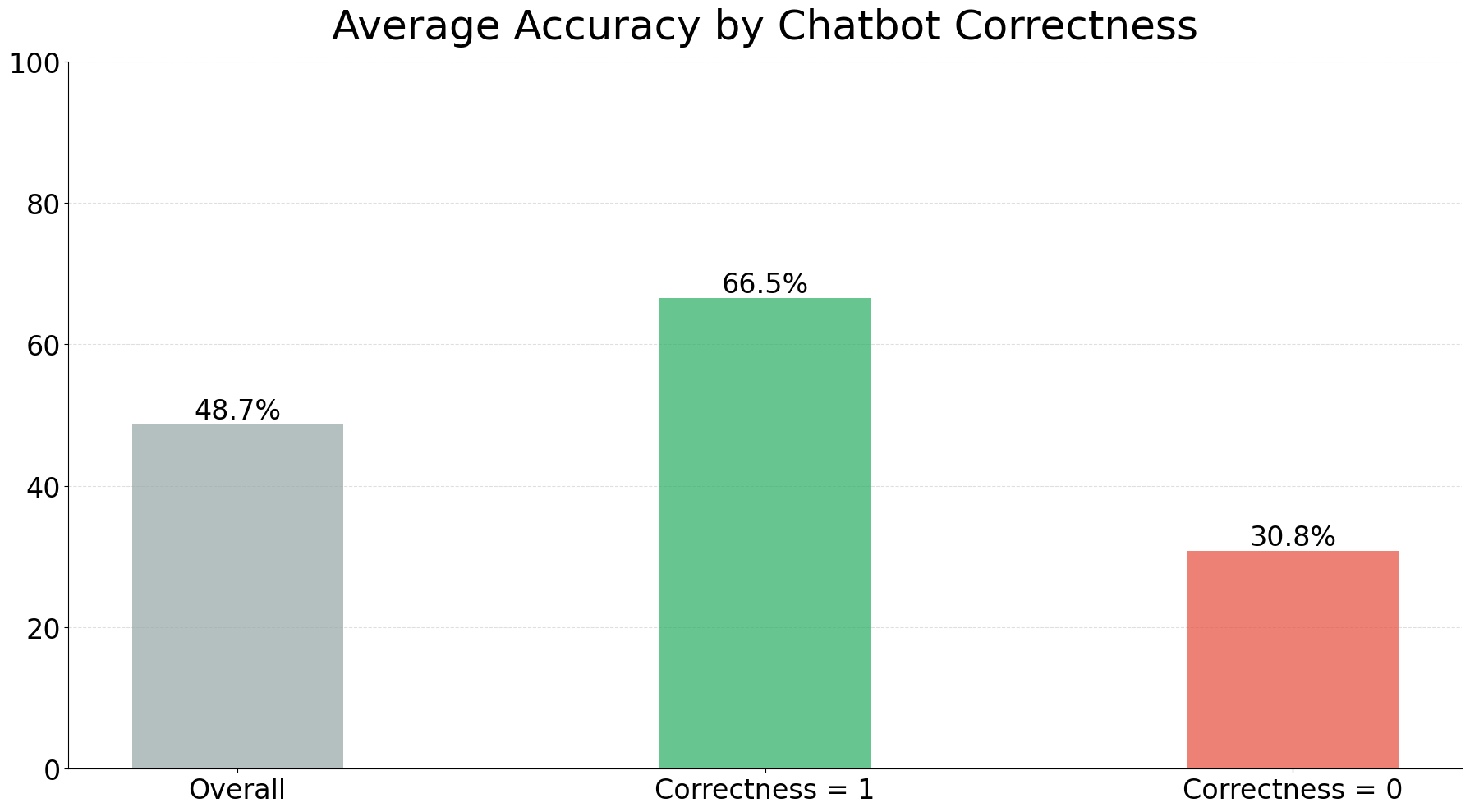}
        \caption{Average accuracy of the participants, depending on the correctness condition.}
        \label{fig:accuracy_per_correctness}
    \end{figure}
        
    To fit a model where accuracy or agreement is the dependent variable, we used GLMMs (Generalized LMM) with logit links (binomial family) for 1194 observations (199 participants $\times$ 6 questions). To better identify participant behavior under different scenarios, we further investigated our dependent variables for correct-only and incorrect-only answers separately. 
    
    For both correctness cases, we had 597 observations. Agreement and accuracy when the chatbot is correct refer to the same conditions; thus, they were analyzed together. Figure~\ref{fig:accuracy_per_correctness} shows the overall accuracy of the participants, based on the correctness condition. To assess model fits, we obtained Pseudo-$R^2$ scores using MuMIn\footnote{\url{https://www.rdocumentation.org/packages/MuMIn/versions/1.48.11}}. $R^2_{marginal}$ represents the variance explained by fixed effects only, and $R^2_{conditional}$ represents the full model (fixed effects and random effects). 

        \subsubsection{Accuracy and Agreement for Correct Chatbot Answers}
    
        Table~\ref{tab:accuracy_cor_results} shows the accuracy and agreement results when chatbot answers were correct. Both questions and participants show noticeable variance: participants' agreement and accuracy tendencies vary, and questions have different success rates. The influence of random effects is clear in the amount of variance they explain: $R^2_{marginal}=0.284$, $R^2_{conditional}=0.646$.
    
        Trust, Familiarity, and AI-URLsNum have significant positive effects on accuracy, while Difficulty\textsubscript{chatbot} has a large, negative one. Participants who are familiar with the topic and have high trust in chatbots are more likely to agree with chatbot responses. This tendency improves accuracy when responses are correct, particularly when confirmed by another AI source. The alignment between multiple sources of information (the chatbot, another AI model, and personal knowledge) is critical for agreement.
        
        Assessing that the question is challenging for the chatbot leads to disagreement and, therefore, to inaccuracy when the chatbot is correct. Given the large effect size, this suggests that people might not be as good as they expect at identifying whether a chatbot is accurate, and hints at underreliance. On the other hand, trusting chatbots and believing in their capabilities leads to accurate choices when the chatbot is correct.
    
        \begin{table}[t]
            \centering
            \begin{tabular}{lcccc}
            \hline
            \multicolumn{5}{l}{\textit{Random Effects}} \\
            \hline\hline
            PID (Intercept)     & \multicolumn{4}{l}{Variance = 2.03, SD = 1.42} \\
            QID (Intercept)& \multicolumn{4}{l}{Variance = 2.62, SD = 1.62} \\
            \hline\hline
            \textbf{Predictor} & \textbf{Estimate} & \textbf{SE} & \textbf{$z$-value} & \textbf{$p$-value} \\
            \hline
            (Intercept)         & 1.11      & 0.74  & 1.50    & .134    \\
            \textbf{Trust}      & 0.54      & 0.19  & 2.80    & $\mathbf{.005^{**}}$  \\
            ChatbotLiteracy    & $\text{-}0.15$   & 0.18  & $\text{-}0.84$ & .402    \\
            WarmthCondition    & 0.66      & 0.35  & 1.90    & .057 \\
            Confidence          & 0.06      & 0.16  & 0.35    & .726    \\
            \textbf{Difficulty\textsubscript{chatbot}} & $\text{-}1.55$   & 0.23  & $\text{-}6.69$ & $\mathbf{<.001^{***}}$ \\
            \textbf{Familiarity}   & 0.35      & 0.17  & 2.09    & $\mathbf{.037^*}$   \\
            SeenOrder          & $\text{-}0.03$   & 0.15  & $\text{-}0.24$ & .814    \\
            InitCorrectness    & $\text{-}0.20$   & 0.35  & $\text{-}0.57$ & .572    \\
            InteractionLen     & 0.17      & 0.16  & 1.08    & .280    \\
            NonAI-URLsNum         & $\text{-}0.14$   & 0.18  & $\text{-}0.79$ & .430    \\
            \textbf{AI-URLsNum }  & 0.53      & 0.23  & 2.34    & $\mathbf{.019^{*}}$   \\
            \hline
            \multicolumn{5}{l}{\footnotesize Note:  *$p < .05$; **$p < .01$; ***$p < .001$} \\
            \end{tabular}
            \caption{GLMM results for participant accuracy (and agreement) when the chatbot's initial answer is correct.}
            \label{tab:accuracy_cor_results}
        \end{table}

        \subsubsection{Accuracy for Incorrect Chatbot Answers}
    
        \begin{table}[t]
            \centering
            \begin{tabular}{lcccc}
            \hline
            \multicolumn{5}{l}{\textit{Random Effects}} \\
            \hline\hline
            PID (Intercept)     & \multicolumn{4}{l}{Variance = 1.04, SD = 1.02} \\
            QID (Intercept)& \multicolumn{4}{l}{Variance = 3.46, SD = 1.86} \\
            \hline\hline
            \textbf{Predictor} & \textbf{Estimate} & \textbf{SE} & \textbf{$z$-value} & \textbf{$p$-value} \\
            \hline
            (Intercept)        & $\text{-}1.75$   & 0.83  & $\text{-}2.11$ & $.035^*$   \\
            \textbf{Trust}               & $\text{-}0.37$   & 0.17  & $\text{-}2.20$ & $\mathbf{.028^*}$   \\
            \textbf{ChatbotLiteracy}    & 0.41      & 0.16  & 2.52    & $\mathbf{.012^*}$   \\
            WarmthCondition             & $\text{-}0.29$   & 0.30  & $\text{-}0.94$ & .347    \\
            \textbf{Confidence}         & 0.38      & 0.15  & 2.52    & $\mathbf{.012^*}$   \\
            \textbf{Difficulty\textsubscript{chatbot}}         & 0.91      & 0.17  & 5.41    & $\mathbf{<.001^{***}}$ \\
            \textbf{Familiarity}        & $\text{-}0.45$   & 0.16  & $\text{-}2.83$ & $\mathbf{.005^{**}}$  \\
            SeenOrder                  & 0.25      & 0.14  & 1.73    & .084    \\
            InitCorrectness            & 0.34      & 0.32  & 1.06    & .288    \\
            InteractionLen             & 0.22      & 0.14  & 1.59    & .112    \\
            NonAI-URLsNum              & 0.08      & 0.14  & 0.55    & .585    \\
            AI-URLsNum                 & 0.11      & 0.11  & 0.98    & .329    \\
            \hline
            \multicolumn{5}{l}{\footnotesize Note:  *$p < .05$; **$p < .01$; ***$p < .001$} \\
            \end{tabular}
            \caption{GLMM results for participant accuracy when the chatbot's initial answer is incorrect.}        
            \label{tab:accuracy_incor_results}
        \end{table}
        
        Accuracy results for incorrect-only chatbot answers are shown in Table~\ref{tab:accuracy_incor_results}. Participants continue to vary, while the variance is smaller than the correct-only answers. Question variance, however, is higher. We believe this is due to the low accuracy in Q3, as discussed in \ref{question_pitfalls} and \ref{question_statistics}. $R^2_{marginal}=0.168$ and $R^2_{conditional}=0.576$ again confirms the influence of random effects.
    
        Compared to the correct-only chatbot answer results, signs of some of the fixed effects change: Trust and Familiarity have significantly negative influences, while Difficulty\textsubscript{chatbot} has a positive one. Confidence and ChatbotLiteracy also have significant, moderate positive effects. Familiarity does not seem to be properly assessed by participants, as it hurts accuracy. Confidence, on the other hand, is assessed better.  
        
        While correct-only answers rewarded trust in chatbots and revealed underreliance, incorrect-only answers show the other side of the coin, as trust becomes a negative influence. This is a clear sign of overreliance, as some participants seem to agree with the chatbot, regardless of the circumstances. To find the correct answer when a chatbot is misleading, being skeptical about its capabilities, and being digitally literate helps.

        \subsubsection{Accuracy for All Answers}

       \begin{table}[t]
            \centering
            \begin{tabular}{lcccc}
            \hline
            \multicolumn{5}{l}{\textit{Random Effects}} \\
            \hline\hline
            PID (Intercept)     & \multicolumn{4}{l}{Variance = 0.00, SD = 0.00} \\
            QID (Intercept)& \multicolumn{4}{l}{Variance = 1.70, SD = 1.30} \\
            \hline\hline
            \textbf{Predictor} & \textbf{Estimate} & \textbf{SE} & \textbf{$z$-value} & \textbf{$p$-value} \\
            \hline
            (Intercept)       & $\text{-}1.21$   & 0.56  & $\text{\text{-}}0.18$ & $.029^*$   \\
            Trust               & 0.06      & 0.08  & 0.75    & .453    \\
            ChatbotLiteracy    & 0.14      & 0.08  & 1.85    & .064 \\
            WarmthCondition    & 0.18      & 0.15  & 1.21    & .225    \\
            Confidence          & 0.10      & 0.08  & 1.33    & .183    \\
            \textbf{Difficulty\textsubscript{chatbot}}          & $\text{-}0.17$   & 0.08  & $\text{-}2.19$ & $\mathbf{.029^*}$   \\
            Familiarity         & $\text{-}0.09$   & 0.07  & $\text{-}1.19 $ & .236    \\
            \textbf{Correctness}         & 2.06      & 0.16  & 12.69   & $\mathbf{<.001^{***}}$ \\
            SeenOrder          & 0.10      & 0.07  & 1.34    & .181    \\
            InitCorrectness    & 0.12      & 0.15  & 0.80    & .427    \\
            \textbf{InteractionLen}     & 0.20      & 0.07  & 2.76    & $\mathbf{.006^{**}}$  \\
            NonAI-URLsNum       & $\text{-}0.03$   & 0.08  & $\text{-}0.44$ & .663    \\
            \textbf{AI-URLsNum}          & 0.20      & 0.08  & 2.66    & $\mathbf{.008^{**}}$  \\
            \hline
            \multicolumn{5}{l}{\footnotesize Note:  *$p < .05$; **$p < .01$; ***$p < .001$} \\
            \end{tabular}
            \caption{GLMM results for participant accuracy.}
            \label{tab:accuracy_results}
        \end{table}
        
        Table~\ref{tab:accuracy_results} shows the combined accuracy results when Correctness is included as a control variable. Looking first at the random effects, we see that PID has zero variance. This means that participant-level variation was practically non-existent. Questions showed variance, but less than in the correct-only and incorrect-only models. We believe this is due to the domineering influence of Correctness, explaining a remarkable proportion of the variance. Pseudo-$R^2$ scores for the model were $R^2_{marginal}=0.188$ and $R^2_{conditional}=0.430$.
    
        Correctness has a large, positive effect: receiving the correct answer from the chatbot significantly helps accuracy. Besides Correctness, InteractionLen and AI-URLsNum also have significant positive effects, but NonAI-URLsNum has almost none. Participant accuracy increases upon receiving help from AI (either from the embedded chatbot or another source), but not from other sources. 
    
        We notice that some significant variables from the previous models were missing in the full model. For Trust and Familiarity, this occurred because their opposing effects on correct versus incorrect responses cancelled each other out when both conditions were combined. The same can be said about Difficulty\textsubscript{chatbot} even though it is significant: its effect is much lower because of opposing directions. ChatbotLiteracy was significant for the incorrect-only model with a positive effect. However, it did not show any significant effect on the correct-only model, reducing its overall influence. InteractionLen was insignificant in any of the correctness models, despite having a small positive effect in both. We believe that since some variables lost their influence on the combined results, InteractionLen had a more important role in explaining the variance.

        \subsubsection{Agreement for Incorrect Chatbot Answers}
    
        \begin{table}[t]
            \centering
            \begin{tabular}{lcccc}
            \hline
            \multicolumn{5}{l}{\textit{Random Effects}} \\
            \hline\hline
            PID (Intercept)     & \multicolumn{4}{l}{Variance = 2.02, SD = 1.42} \\
            QID (Intercept)& \multicolumn{4}{l}{Variance = 0.40, SD = 0.63} \\
            \hline\hline
            \textbf{Predictor} & \textbf{Estimate} & \textbf{SE} & \textbf{$z$-value} & \textbf{$p$-value} \\
            \hline
            (Intercept)        & $\text{-}0.60$   & 0.40  & $\text{-}1.52$ & .129   \\
            \textbf{Trust}              & 0.71      & 0.19  & 3.82    & $\mathbf{<.001^{***}}$    \\
            ChatbotLiteracy    & $\text{-}0.15$      & 0.17  & $\text{-}0.89$    & .372 \\
            \textbf{WarmthCondition}    & 0.72      & 0.33  & 2.17    & $\mathbf{.030^*}$    \\
            \textbf{Confidence}          & $\text{-}0.29$      & 0.15  & $\text{-}2.00$    & $\mathbf{.046^*}$    \\
            \textbf{Difficulty\textsubscript{chatbot}}          & $\text{-}1.87$   & 0.22  & $\text{-}8.50$ & $\mathbf{<.001^{***}}$   \\
            \textbf{Familiarity}         & 0.42   & 0.16  & 2.66 & $\mathbf{.008^{**}}$    \\
            SeenOrder          & $\text{-}0.21$      & 0.14  & $\text{-}1.49$    & .137    \\
            InitCorrectness    & $\text{-}0.12$      & 0.34  & $\text{-}0.35$    & .728    \\
            InteractionLen     & 0.01      & 0.16  & 0.09    & .930  \\
            NonAI-URLsNum       & 0.11   & 0.16  & 0.68 & .499    \\
            AI-URLsNum          & $\text{-}0.16$      & 0.17  & $\text{-}0.90$    & .368  \\
            \hline
            \multicolumn{5}{l}{\footnotesize Note:  *$p < .05$; **$p < .01$; ***$p < .001$} \\
            \end{tabular}
            \caption{GLMM results for agreement when the chatbot’s initial answer is incorrect.}
            \label{tab:agreement_incor_results}
        \end{table}
    
        Table~\ref{tab:agreement_incor_results} shows the agreement results when the chatbot's initial answer is incorrect. Participant-level variance is high (2.02), while question-level variance is moderate (0.40). The explained variance shows that both fixed and random effects account for a substantial portion of the variance ($R^2_{marginal}=0.431$ and $R^2_{conditional}=0.643$).

        Trust and Difficulty\textsubscript{chatbot} are significant predictors: Trust in chatbots increases the tendency to agree, while assessing the question as too difficult for the chatbot to answer significantly reduces agreement. Additionally, three other variables reach significance: Familiarity increases agreement, while Confidence decreases it. Notably, WarmthCondition also reaches significance: participants agree more with the warm chatbot when the answer is incorrect. 
        
        \subsubsection{Agreement for All Answers}
        
        \begin{table}[t]
            \centering
            \begin{tabular}{lcccc}
            \hline
            \multicolumn{5}{l}{\textit{Random Effects}} \\
            \hline\hline
            PID (Intercept)     & \multicolumn{4}{l}{Variance = 2.00, SD = 1.41} \\
            QID (Intercept)& \multicolumn{4}{l}{Variance = 0.09, SD = 0.30} \\
            \hline\hline
            \textbf{Predictor} & \textbf{Estimate} & \textbf{SE} & \textbf{$z$-value} & \textbf{$p$-value} \\
            \hline
            (Intercept)        & $\text{-}0.26$   & 0.28  & $\text{-}0.94$ & .350   \\
            \textbf{Trust}              & 0.51      & 0.14  & 3.56    & $\mathbf{<.001^{***}}$    \\
            ChatbotLiteracy    & $\text{-}0.11$      & 0.14  & $\text{-}0.81$    & .416 \\
            WarmthCondition    & 0.47      & 0.27  & 1.77    & .077    \\
            Confidence          & 0.00      & 0.11  & 0.02    & .985    \\
            \textbf{Difficulty\textsubscript{chatbot}}          & $\text{-}1.83$   & 0.14  & $\text{-}13.27$ & $\mathbf{<.001^{***}}$   \\
            Familiarity         & 0.19   & 0.12  & 1.65 & .100    \\
            \textbf{Correctness}         & 1.12      & 0.17  & 6.42   & $\mathbf{<.001^{***}}$ \\
            SeenOrder          & $\text{-}0.11$      & 0.09  & $\text{-}1.24$    & .216    \\
            InitCorrectness    & $\text{-}0.27$      & 0.27  & $\text{-}1.01$    & .311    \\
            InteractionLen     & $\text{-}0.00$      & 0.11  & $\text{-}0.02$    & .985  \\
            NonAI-URLsNum       & 0.07   & 0.12  & 0.63 & .530    \\
            AI-URLsNum          & 0.05      & 0.11  & 0.42    & .676  \\
            \hline
            \multicolumn{5}{l}{\footnotesize Note:  *$p < .05$; **$p < .01$; ***$p < .001$} \\
            \end{tabular}
            \caption{GLMM results for agreement.}
            \label{tab:agreement_results}
        \end{table}

        Agreement for all answers follows a similar pattern to the incorrect-only model (Table~\ref{tab:agreement_results}). Participant variance remains high (2.00). Question-level variance is considerably lower than in the incorrect-only model (0.09), as supported by Figure~\ref{fig:agreement}. The explained variance is also comparable to the incorrect-only model ($R^2_{marginal}=0.433$ and $R^2_{conditional}=0.626$).

        Contrary to the full accuracy model, Correctness does not have the largest effect size, despite being a significant predictor. It is Difficulty\textsubscript{chatbot} that surpasses even Correctness: It significantly reduces agreement regardless of correctness conditions. Trust is a significant positive predictor of agreement, again, regardless of the correctness. While Familiarity has a positive influence on both correct and incorrect only models, it does not reach significance for the full model due to the small effect size. Confidence has a small negative effect on incorrect answers and is ineffective for correct answers. For the full model, it does not influence the outcome. WarmthCondition is a positive predictor for both correctness models, even though it is significant only for incorrect answers. It does not reach significance in the full model ($p = .077$). Finally, AI-URLsNum is also not significant; it has a significant positive effect on correct-only answers, but a negative influence on incorrect-only answers.
                
        \begin{figure}[t]
            \centering
            \includegraphics[width=\linewidth]{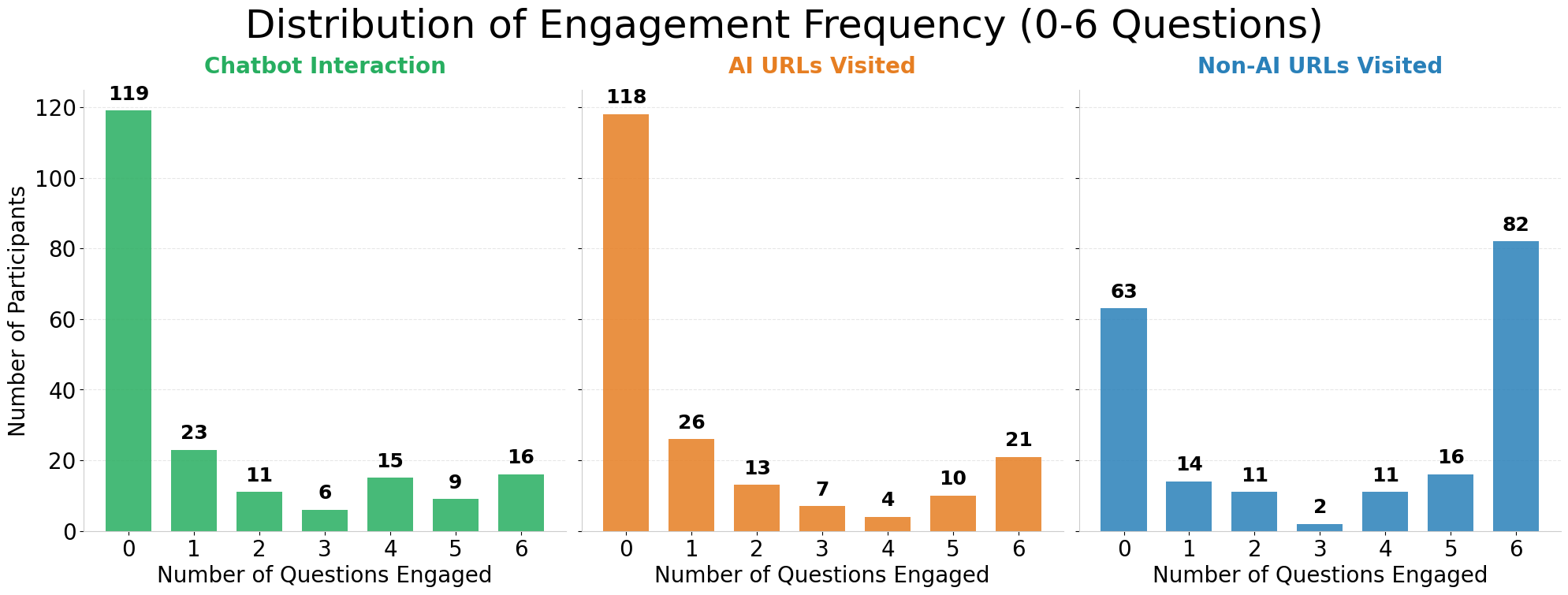}
            \caption{Histograms showing the distribution of usage frequency across three behavioral variables. The x-axis represents the number of questions for which a participant actively engaged with the specific tool. The y-axis represents the count of participants falling into each frequency bucket.}
            \label{fig:interaction_distribution}
        \end{figure}
    
    \subsection{Behavioral Analysis}
    
    In this section, we examine behavioral variables, focusing specifically on how fact-checking behavior influences reliance and what drives the decision to fact-check.

        \subsubsection{Engagement Decision}
        
        Figure~\ref{fig:interaction_distribution} shows the distribution of the number of questions participants engaged in fact-checking behavior. InteractionLen and AI-URLsNum follow left-skewed distributions with zero being the dominant value. NonAI-URLsNum, however, has a U-shaped distribution: the majority of participants checked NonAI sources for all six questions (n=82) or not even once (n=63). This finding shows that fact-checking behavior is driven by personal choice. While some users felt the need to fact-check constantly, others found it unnecessary for any question. \ref{descriptive_statistiscs} provides detailed descriptive statistics for behavioral variables.
    
        Engagement was not mutually exclusive in our experimental design; participants were free to interact with all three sources of information (chatbot, AI URLs, and Non-AI URLs). Taking this into account, we see that only 13\% of the participants did not engage with any source throughout the task (see Table~\ref{tab:behavioral_desc}). Figure~\ref{fig:interaction_distribution} shows that Non-AI URLs are the most common sources of fact-checking. Chatbot engagement and AI URLs show similar distributions; even the number of participants who did not engage at all is almost the same. We believe this suggests that using an AI solution is also a personal choice, independent of the context.
        
        \begin{table}[htbp]
            \centering
            \begin{tabular}{lcccc}
            \hline
            \multicolumn{5}{l}{\textit{Random Effects}} \\
            \hline\hline
            PID (Intercept)     & \multicolumn{4}{l}{Variance = 21.10, SD = 4.59} \\
            Question (Intercept)& \multicolumn{4}{l}{Variance = 00.00, SD = 0.00} \\
            \hline\hline
            \textbf{Predictor} & \textbf{Estimate} & \textbf{SE} & \textbf{$z$-value} & \textbf{$p$-value} \\
            \hline
            (Intercept)        & $\text{-}4.54$   & 0.78  & $\text{-}5.79$ & $<.001^{***}$   \\
            \textbf{Trust}              & 0.99      & 0.46  & 2.14    & $\mathbf{.032^{*}}$    \\
            ChatbotLiteracy    & 0.35      & 0.44  & 0.80    & .425 \\
            WarmthCondition    & $\text{-}0.72$      & 0.84  & $\text{-}0.86$    & .388    \\
            Difficulty\textsubscript{chatbot}       & 0.23      & 0.14  & 1.65    & .100    \\
            Familiarity         & $\text{-}0.11$   & 0.18  & $\text{-}0.62$ & .534   \\
            Correctness         & $\text{-}0.20$   & 0.24  & $\text{-}0.83$ & .408    \\
            \textbf{SeenOrder}          & $\text{-}0.44$      & 0.12  & $\text{-}3.71$    & $\mathbf{<.001^{***}}$  \\
            InitCorrectness    & 1.07      & 0.85  & 1.26    & .209    \\
            \hline
            \multicolumn{5}{l}{\footnotesize Note:  *$p < .05$; **$p < .01$; ***$p < .001$} \\
            \end{tabular}
            \caption{GLMM results for AnyInteractions, a binary indicator of whether the participant interacted with the chatbot.}
            \label{tab:any_interactions_results}
        \end{table}
    
       \begin{table}[t]
            \centering
            \begin{tabular}{lcccc}
            \hline
            \multicolumn{5}{l}{\textit{Random Effects}} \\
            \hline\hline
            PID (Intercept)     & \multicolumn{4}{l}{Variance = 22.87, SD = 4.78} \\
            Question (Intercept)& \multicolumn{4}{l}{Variance = 00.01, SD = 0.09} \\
            \hline\hline
            \textbf{Predictor} & \textbf{Estimate} & \textbf{SE} & \textbf{$z$-value} & \textbf{$p$-value} \\
            \hline
            (Intercept)        & $\text{-}4.86$   & 1.00  & $\text{-}4.86$ & $<.001^{***}$   \\
            \textbf{Trust}              & $\text{-}1.18$      & 0.40  & $\text{-}2.99$    & $\mathbf{.003^{**}}$    \\
            ChatbotLiteracy    & 0.39      & 0.39  & 1.00    & .317 \\
            WarmthCondition    & 0.24      & 0.77  & 0.32    & .751    \\
            Difficulty\textsubscript{chatbot}   & 0.20      & 0.14  & 1.41    & .159    \\
            Familiarity         & $\text{-}0.28$   & 0.21  & $\text{-}1.30$ & .193   \\
            Correctness         & $\text{-}0.13$   & 0.24  & $\text{-}0.54$ & .592    \\
            SeenOrder          & 0.04      & 0.12  & 0.30    & .762    \\
            InitCorrectness    & 0.56      & 0.77  & 0.73    & .465    \\
            \hline
            \multicolumn{5}{l}{\footnotesize Note:  *$p < .05$; **$p < .01$; ***$p < .001$} \\
            \end{tabular}
            \caption{GLMM results for AnyAI-URLs, a binary indicator of whether the participant visited any AI-related URL.}
            \label{tab:any_ai_urls_results}
        \end{table}
        
        \begin{table}[htbp]
            \centering
            \begin{tabular}{lcccc}
            \hline
            \multicolumn{5}{l}{\textit{Random Effects}} \\
            \hline\hline
            PID (Intercept)     & \multicolumn{4}{l}{Variance = 28.50, SD = 5.33} \\
            Question (Intercept)& \multicolumn{4}{l}{Variance = 00.00, SD = 0.00} \\
            \hline\hline
            \textbf{Predictor} & \textbf{Estimate} & \textbf{SE} & \textbf{$z$-value} & \textbf{$p$-value} \\
            \hline
            (Intercept)        & 0.46   & 0.80  & 0.57 & .567   \\
            \textbf{Trust}              & $\text{-}2.72$      & 0.50  & $\text{-}5.46$    & $\mathbf{<.001^{***}}$    \\
            ChatbotLiteracy    & 0.58      & 0.46  & 1.28    & .201 \\
            WarmthCondition    & $\text{-}0.59$      & 0.88  & $\text{-}0.67$    & .502    \\
            \textbf{Difficulty\textsubscript{chatbot}}       & 0.41      & 0.15  & 2.69    & $\mathbf{.007^{**}}$    \\
            Familiarity         & $\text{-}0.29$   & 0.18  & $\text{-}1.57$ & .116   \\
            Correctness         & 0.02   & 0.25  & 0.10 & .922    \\
            SeenOrder          & 0.07      & 0.12  & 0.58    & .560    \\
            InitCorrectness    & 1.54      & 0.88  & 1.75    & .080    \\
            \hline
            \multicolumn{5}{l}{\footnotesize Note:  *$p < .05$; **$p < .01$; ***$p < .001$} \\
            \end{tabular}
            \caption{GLMM results for AnyNonAI-URLs, a binary indicator of whether the participant visited any NonAI-related URL.}
            \label{tab:any_nonai_urls_results}
        \end{table}    
       
        \subsubsection{Engagement Prediction}
        \label{engagement_prediction}
    
        To confirm that engagement is a personal choice instead of a contextual one, we fit GLMMs to the behavioral variables. Given the zero inflation, especially in InteractionLen and AI-URLsNum, and our desire to predict behavior, we discretized the variables into a binary choice: counts above zero were labeled ``1: engaged'' and zero counts were labeled ``0: not-engaged''. We named the new behavioral variables AnyInteractions, AnyAI-URLs, and AnyNonAI-URLs. We did not update the accuracy and agreement models presented in the previous section with discrete behavioral variables because they produced identical results as their continuous counterparts. 
    
        Table~\ref{tab:any_interactions_results}, Table~\ref{tab:any_ai_urls_results}, and Table~\ref{tab:any_nonai_urls_results} show the results for AnyInteractions, AnyAI-URLs, and AnyNonAI-URLs. As predictors, we used the same set of variables as before (except for the behavioral variables). For all three models, participant variance is extremely high while question variance is nonexistent. This confirms that fact-checking behavior is a participant-related choice independent of the question. Trust influences AnyInteractions positively and AnyAI-URLs and AnyNonAI-URLs negatively: Participants with high trust in chatbots preferred the experiment chatbot over other AI or Non-AI sources.
            
        Besides Trust, SeenOrder is significant for AnyInteractions: the later a question is encountered, the lower the likelihood of interacting with the chatbot. We believe this was due to survey fatigue. For AnyAI-URLs, no other fixed effect besides trust is significant. Difficulty\textsubscript{chatbot} is a positive predictor of AnyNonAI-URLs: participants searched for Non-AI sources if they believed the question was too challenging for the chatbot.
        
        \subsubsection{Engagement's Role in Accuracy}
    
        We have identified that engagement behavior is highly personal, and engaging with the chatbot or another AI tool increases accuracy. Participant variation for accuracy, on the other hand, was zero when all samples were used (see Table~\ref{tab:accuracy_results}). This raises an important question about the impact of engagement on accuracy: is it due to between-participant or within-participant variation? 
    
        \begin{itemize}
            \item Do people who generally interact more have higher overall accuracy? (Between)
            \item On trials where a participant interacts more than their personal average, are they more accurate than their personal average? (Within)
        \end{itemize}
    
        To answer these questions, we created between and within-behavioral variables for the accuracy model. For brevity, we present only the model with all samples (both correct and incorrect). According to Table~\ref{tab:accuracy_behavioral_results}, AI-URLsNum\textsubscript{between} is highly significant ($p < .001$) while AI-URLsNum\textsubscript{within} has zero influence. This clarifies that the tendency to seek out AI sources predicts accuracy, independent of how much participants interact with the source. For InteractionLen, only the within-variant reaches significance, but the effect sizes are similar. 
    
        \begin{table}[t]
            \centering
            \begin{tabular}{lcccc}
            \hline
            \multicolumn{5}{l}{\textit{Random Effects}} \\
            \hline\hline
            PID (Intercept)     & \multicolumn{4}{l}{Variance = 0.00, SD = 0.00} \\
            Question (Intercept)& \multicolumn{4}{l}{Variance = 1.72, SD = 1.31} \\
            \hline\hline
            \textbf{Predictor} & \textbf{Estimate} & \textbf{SE} & \textbf{$z$-value} & \textbf{$p$-value} \\
            \hline
            (Intercept)        & $\text{-}1.20$   & 0.56  & $\text{-}2.15$ & $.032^{*}$   \\
            Trust              & 0.09      & 0.08  & 1.06    & .290    \\
            ChatbotLiteracy    & 0.12      & 0.08  & 1.53    & .126 \\
            WarmthCondition    & 0.17      & 0.15  & 1.14    & .254    \\
            Confidence         & 0.09      & 0.08  & 1.21    & .226    \\
            \textbf{Difficulty\textsubscript{chatbot}}          & $\text{-}0.16$      & 0.08  & $\text{-}2.12$    & $\mathbf{.034^{*}}$    \\
            Familiarity         & $\text{-}0.07$   & 0.08  & $\text{-}0.99$ & .322   \\
            \textbf{Correctness}         & 2.08   & 0.16  & 12.69 & $\mathbf{<.001^{***}}$    \\
            SeenOrder          & 0.08      & 0.07  & 1.09    & .274    \\
            InitCorrectness    & 0.10      & 0.15  & 0.66    & .511    \\
            InteractionLen\textsubscript{between}  & 0.13      & 0.08  & 1.74    & .082    \\
            \textbf{InteractionLen\textsubscript{within}}   & 0.15      & 0.07  & 2.09    & $\mathbf{.037^{*}}$    \\
            \textbf{AI-URLsNum\textsubscript{between}}       & 0.29   & 0.08  & 3.56 & $\mathbf{<.001^{***}}$    \\
            AI-URLsNum\textsubscript{within}        & 0.00      & 0.07  & 0.03    & .978  \\
            NonAI-URLsNum\textsubscript{between}   & $\text{-}0.00$   & 0.08  & $\text{-}0.04$ & .969    \\
            NonAI-URLsNum\textsubscript{within}    & $\text{-}0.03$      & 0.07  & $\text{-}0.41$    & .683  \\
            \hline
            \multicolumn{5}{l}{\footnotesize Note:  *$p < .05$; **$p < .01$; ***$p < .001$} \\
            \end{tabular}
            \caption{GLMM results including between- and within-participant behavioral variables for accuracy.}
            \label{tab:accuracy_behavioral_results}
        \end{table}

    \subsection{Qualitative Analysis}

    After completing the task, participants answered a mandatory open-ended question about how they decided whether to trust or fact-check the chatbot's responses. We analyzed these answers using thematic coding, labeling responses based on three key themes: verification behavior, default trust in chatbots, and perceived warmth.
    
        \subsubsection{Verification Behavior}
    
        Participants are categorized into three groups based on their verification approach: a small group (n=33) that always fact-checks answers, another group (n=51) that did not mention anything about fact-checking, and the largest group (n=115) representing conditional fact-checkers: participants who check other sources based on certain criteria. The three distinct groups align with our behavioral finding that the fact-checking approach is rooted in participant characteristics.
        
        Most participants mentioned intuition as a driver of fact-checking (\textit{the answer sounded wrong}); however, they were clearer about the driver of picking an answer: source consistency. The most common theme in open-ended answers was agreement between multiple sources. Given our setup, the chatbot was the first source of information for participants. The second most common source was web search: many participants specifically mentioned checking web search results to verify the chatbot's answer. Furthermore, they mentioned preferring search results if there was a disagreement between sources (\textit{If multiple web search results led to the same answer, I favored it over the bot}).
    
        \subsubsection{Default Trust in Chatbots}
    
        While most participants did not mention trust specifically (n=149), some stated that they either trust (n=19) or distrust (n=31) chatbots by default. One of our participants mentioned: \textit{I use chatgpt through my day, so I kind of blindly trust it}, clearly showing overreliance. On the other hand, another participant claimed: \textit{They (chatbots) were mostly wrong, I don't think AI or chatbots are trustworthy}, showing underreliance. 
    
        \subsubsection{Perceived Warmth}
    
        Only a minority of the participants (n=7) explicitly mentioned warmth in their answers. Interestingly, 4 of them mentioned it as a negative (\textit{The chatbot was frequently incorrect despite its very cheerful and enthusiastic demeanor.}) and 3 mentioned it as a positive (\textit{The enthusiasm that the chatbot used when giving me the answers as well as the information it used to back up its sources made me trust it.}). 

\section{Discussion}

    In this section, we evaluate our hypotheses in light of the study findings to address our research questions.
    
    \subsection{Fact-checking as a Solution for Reliance}

    The first hypothesis claimed that being able to fact-check does not eliminate overreliance or underreliance in conversational search. Our findings confirm the existence of both, though they manifest in different ways. 
    
    For some participants, reliance patterns were straightforward to identify. A small subset (25.6\%) explicitly mentioned trusting or distrusting chatbots by default in their open-ended responses, and their behavior aligned with the stated tendencies. However, for the majority of participants, reliance was more indirect and could not be inferred solely from behavior. The vast majority (87\%) interacted with at least one fact-checking tool throughout the experiment, yet this engagement does not necessarily indicate the absence of reliance. As we demonstrate in the following subsections, reliance can persist even when users actively consult additional sources.

        \subsubsection{Manifestations of Overreliance}
        \label{manifestations_overreliance}

        High overall trust in chatbots led some participants to accept responses without verification, even when fact-checking tools were available. Furthermore, it influenced behavior: high-trust participants were eager to interact with the chatbot but not with external sources. This means instead of looking for alignment between multiple sources, they used the chatbot for verification. This is a clear case of overreliance; misaligned trust led participants to either not fact-check at all or use the same chatbot that provided the information as the fact-checking source. 

        Even for low-trust participants, the chatbot's presence shaped their verification behavior. Participants cited alignment between sources as their primary criterion for validation. However, this alignment was anchored to the chatbot's answer, as it was the first information source they encountered. This anchoring effect helps explain why NonAI-URLsNum was not a positive predictor of accuracy, despite being the most common behavioral variable (Figure~\ref{fig:interaction_distribution}). Participants treated a single corroborating source as sufficient validation, aligning with \citet{Kim2024Certainty}'s findings that AI's presence makes verification shallow and does not necessarily improve performance. This represents a form of overreliance.

        A more indirect indicator of overreliance emerged in participants' agreement patterns: they agreed more frequently with the warm chatbot, specifically when it provided incorrect information. Since participants were unaware of the correctness manipulation, they approached the task as selecting the correct answer from conflicting sources. When the chatbot's initial response was incorrect and contradicted information from other sources, participants faced increased uncertainty. Prior research shows that in the presence of AI, uncertainty does not prompt additional information-seeking \citep{Kim2024Certainty}. In this context, agreeing with the chatbot under uncertainty reflects a decision to trust the AI rather than invest effort in further verification. Our findings suggest that conversational warmth nudges users to trust AI in uncertain situations, demonstrating a form of overreliance.
        
        \subsubsection{Manifestations of Underreliance}

        Underreliance is often overlooked compared to overreliance, yet it can be equally harmful \citep{hashemi2024evaluating}. Some participants consulted Non-AI URLs for every question and reported prioritizing web search results over chatbot responses. While web search can be helpful, the assumption that it is always superior to chatbots is problematic. Users tend to prefer search results confirming their existing beliefs \citep{Elsweiler2025query}, and searching can actually increase belief in misinformation \citep{Aslett2024Misinformation}. For participants with low trust in AI, this indiscriminate preference for web search over chatbot responses represents underreliance.
        
        A similar underreliance pattern emerged with Difficulty\textsubscript{chatbot}: when participants judged a question as too difficult for the chatbot to answer, they were less likely to agree with its response. While this skepticism helped participants reject incorrect answers, it also led them to dismiss correct ones. The negative effect of Difficulty\textsubscript{chatbot} on accuracy in the full model, combined with its larger effect size in the correct-only model, demonstrates that participants' underestimation of chatbot capabilities reduced their performance.

        \subsubsection{Practical Guidance}

        In this section, we provide practical implications of our work by outlining an ideal strategy for mitigating both types of reliance on conversational search based on our findings.

        Trust in chatbots, Difficulty\textsubscript{chatbot}, and Familiarity were significant predictors in both correctness models, but with opposite signs, indicating that whether they help depends on chatbot correctness. Users with high trust, those who believe they are familiar with the topic, or those who perceive questions as easy for the chatbot should approach chatbot responses with greater skepticism. Conversely, users with low trust or those who underestimate chatbot capabilities would benefit from moderating their skepticism. Regardless of tendencies, chatbot literacy helps overall with accuracy, especially when uncertainty is high. Given similar previous findings \citep{exploring_motivations, bias_perception}, increasing literacy about common AI terms would be beneficial.
        
        Besides Correctness, AI-URLsNum\textsubscript{between} was the most significant positive influence on accuracy. Alignment between sources is important for users to decide on an answer. However, individual websites can be misleading, and checking multiple websites increases user effort. Since chatbots conduct web searches, they can efficiently synthesize sources. According to our findings, merely consulting other sources combats reliance, and prolonged conversations are not necessary. Finally, users should exercise more caution when chatbot answers are warm, since it subconsciously leads them to agree with the chatbot more.
                       
    \subsection{Origins of Verification Behavior}
    
    What drives users' decisions to fact-check chatbot answers? For all three of our behavioral variables, the most important determinant was the participant's own tendency: Some participants were prone to verify or not verify the answer regardless of the context. Furthermore, participants showed individual differences in source preferences: some consulted other AI sources while others used non-AI sources exclusively. Therefore, the decision to verify a chatbot’s answer is primarily driven by user-level characteristics, which aligns with the second hypothesis.
        
    Existing trust in chatbots was the most significant predictor of verification behavior between fixed-effects, supporting our hypothesis. While trust increased the likelihood of interacting with the experiment chatbot, it significantly reduced the probability of verifying the answer with other sources, with the effect being more pronounced for Non-AI URLs. This shows that high-trust users do not prefer to fact-check with external sources, but if they do, they have a more positive attitude towards AI sources.

    Difficulty\textsubscript{chatbot} was a significant predictor of visiting any Non-AI URL, but was insignificant for other behavioral variables. We believe this reflects the difference in attitude towards AI: participants who thought the question was too difficult for the chatbot used Non-AI sources for verification. This aligns with the finding that the preferred source type (AI vs. Non-AI) is also a personal decision: Some participants believed questions are difficult for AI systems in general, not only the experiment chatbot.

    Interestingly, ChatbotLiteracy did not predict any verification behaviors, yet it improved accuracy when the chatbot provided incorrect answers. This suggests that chatbot literacy influences not whether users verify, but how they evaluate information during verification. Individuals with lower digital literacy are more likely to encounter and accept unreliable web search results \citep{Aslett2024Misinformation}. Similarly, chatbot literacy may enhance users' ability to critically assess the quality and credibility of both chatbot responses and external sources they consult, rather than simply increasing verification frequency.
                
    \subsection{Influence of Warmth on Reliance}

    Warmth did not predict any verification behavior, nor did it significantly influence accuracy. Furthermore, only 7 participants mentioned warmth in the open-ended answers. As a direct effect on accuracy, conversation warmth looks unpromising. However, it showed a crucially significant indirect effect: participants agreed more with the warm chatbot when the chatbot's answer was incorrect, as discussed in Section~\ref{manifestations_overreliance}. 

    Considering that most participants did not mention warmth, we believe it works as an indirect mechanism. The vast majority of participants employed conscious ways to curb overreliance, such as fact-checking. Still, they had to decide on an answer given multiple, sometimes conflicting sources. Warmth played a subconscious role here by nudging participants to agree with the chatbot's answer, without participants noticing. This aligns with \citet{financial_advisors}'s finding that users emotionally trust warm LLMs more, despite receiving worse guidance.

    According to the third hypothesis: \textit{A warm conversation style increases user agreement with chatbot answers, thereby amplifying overreliance}. While warmth increased agreement in both correctness conditions, it only became significant for incorrect answers. Therefore, we partially accept our hypothesis. However, we believe this is an even more consequential finding; agreeing with chatbots when they are misleading can have serious implications. Thus, conversation warmth should be taken into account while studying overreliance.

    Of the 7 participants who mentioned warmth in their answers, there was no consensus: some found it to be a positive effect, while others believed it reduced credibility. The small sample size does not allow us to explore how warmth might be perceived differently, but suggests that the influence of warmth might be personal. 

\section{Limitations and Future Work}

There are several limitations of our work. We believe some of the limitations are due to the browser extension: Most people abandoned the survey after the introduction due to this constraint. While our participant group is demographically diverse, people who agree to install an extension might not be representative of the general population. Furthermore, tracking may not have been completely accurate. It is possible that some participants completed the survey on a separate browser window to avoid getting tracked. For future work, a laboratory setting can be considered.

We aimed the experiment to be realistic, hence, our choice to allow participants to access any source freely while keeping it as controlled as possible. However, having access to AI tools risks participants encountering different answers, especially from search engine AI-summaries and chatbots, since they are not deterministic. On top of that, AI tools, especially AI-summaries, are subject to frequent changes. While we tracked participant movements, we did not track the answers they encountered or their interactions within websites. A laboratory setting with a predetermined, static set of sources would make the experiment more controlled, despite being less realistic. Future work should explore the balance between realism and control to see if our results can be validated.

Following the previous point, future research can conduct a more granular analysis of verification behavior. A key strength of our study is its nuanced approach to fact-checking. Rather than treating verification as a binary choice, we analyzed source usage through both frequency (number of visits) and type (AI versus Non-AI sources). Future studies could extend this framework by incorporating credibility assessments of consulted sources. Additionally, researchers could examine interaction patterns to determine how the order in which sources are encountered influences decision-making.

The number of observations (participant $\times$ questions) was chosen to balance statistical power and experiment costs. Given the time participants spent in the survey, we believe the number of questions was balanced. However, since we crafted the questions ourselves, future work can explore a larger, more diverse set.

According to our results, random effects play a large role in explaining the variance for both accuracy prediction and behavioral analysis. While this identifies that overreliance and underreliance operate on a personal level, it does not pinpoint the causes of reliance tendencies. Trust in chatbots was significant in answering our research questions, but it also does not explain the root causes. Future work should identify the individual-level factors that shape reliance behavior and trust formation in chatbot interactions.

\section{Conclusion}

In this paper, we contribute to the literature by showing that reliance manifests at the individual level, independent of AI accuracy or access to verification tools. The decision to fact-check and the manner of fact-checking vary considerably across users: Some trust chatbots by default and rarely verify responses, while others remain skeptical regardless of circumstances. Beyond these individual differences, we identify two key predictors of reliance behavior: prior trust in chatbots and perceived difficulty of the question for the chatbot.

We further contribute by revealing how conversational warmth subconsciously influences decision-making. Warm conversational style nudges users to agree with chatbot responses, particularly when uncertainty is high due to conflicting information. Finally, we introduce chatbot literacy as a protective factor against over-reliance: users with higher chatbot literacy are better able to identify accurate answers when AI provides misleading information.

With this work, we demonstrate evidence that the new paradigm of information search (users accessing conversational and traditional search simultaneously) is insufficient to eliminate reliance. While being skeptical of AI and relying solely on web search for fact-checking alleviates overreliance, it is not necessarily an ideal strategy. Because of confirmation bias and the necessity to check multiple sources for challenging questions, user search inquiries are not guaranteed to be accurate. Our results show that verification with other AI tools can be an alternative strategy. We believe our findings can motivate researchers to investigate the root causes of reliance and identify what makes it personal. As for companies and designers, our work can help them build trustworthy conversational AI agents.

\section*{Acknowledgments}
\noindent
This research is part of the [Anonymized] project with number NWA.1389.20.183 of the research program NWA\_ORC 2020\/21.

\section*{Funding}
\noindent
This research did not receive any specific grant from funding agencies in the public, commercial, or not-for-profit sectors.

\section*{Declaration of Interest}
\noindent
The authors of this manuscript have no potential conflicts of interest to disclose.


    
    
\appendix
\newpage

\section{Common Question Pitfalls}
\label{question_pitfalls}

The reason why chatbots, as well as AI-generated search summaries, are prone to failing the questions is explained as follows:

\begin{itemize}
    \item \textbf{Q1:} \textit{Which global investment firm recently committed a \$5 billion India-focused fund in 2025?} No company has made such an investment, yet, the existence of multiple companies making similar investments (in previous years, in different amounts, or as multiple smaller investments that do not add up to \$5 billion) tends to confuse chatbots and AI-generated search summaries.
    \item \textbf{Q2:} \textit{Which country has the world’s largest forested area per capita?} French Guiana has the largest forested area per capita, but it is a territory. Chatbots sometimes omit this information, leading to inaccurate responses. We also noticed that AI-generated search summaries tend to bring up Canada or Russia. Both countries have substantial forested areas, but looking at per capita, Suriname was in the lead during the experimentation period. As of November 29, Guyana surpassed Suriname\footnote{\url{https://www.visualcapitalist.com/mapped-countries-with-the-most-forest-area-per-capita/}}. Since our experiment was concluded before the 29th, we have treated Suriname as the correct answer.
    \item \textbf{Q3:} \textit{In what year did a country first recognize internet access as a basic human right?} Estonia was the first country to establish a legal basis for internet access in 2000, but it was not framed as a recognition of a basic human right. The lack of attention to the wording sometimes leads chatbots to suggest the year 2000. While preparing the question, we noticed that Q3 had the highest disagreement between sources. It is also the hardest question for the chatbots to answer, and it has the lowest participant accuracy (see Figure~\ref{fig:accuracy}). The framing of ``basic human right'' makes it challenging, since it requires familiarity with legal terminology for clarification. If we investigate the first country that \textit{declared} the internet as a human right, Estonia in 2000 is the correct answer. However, official recognition as a \textit{basic human right} happened first in France in 2009.
    \item \textbf{Q4:} \textit{Which non-primate animals were the first to show mirror self-recognition?} The confusion here is rooted in the existence of multiple studies. In 2001, bottlenose dolphins became the first animals to show mirror self-recognition. A similar study in 2006 recognized the same for Asian elephants. The latter became more famous, and multiple sources claimed this was the first instance, leading chatbots to fail. Sometimes, AI-generated summaries claim Eurasian magpies to be the first, according to a study from 2008. This is supported by some Reddit posts, as well as Wikipedia\footnote{\url{https://en.wikipedia.org/wiki/Mirror_test}}, which we believe is the reason behind the confusion.
    \item \textbf{Q5:} \textit{Which South American country has the second-highest plant biodiversity after Brazil?} There is a mismatch between sources about the number of plant species in Venezuela. While the most recent statistics from multiple sources show Colombia to have a higher biodiversity, the existence of other sources tends to confuse chatbots. Even in the Wikipedia entry of ``Megadiverse countries\footnote{\url{https://en.wikipedia.org/wiki/Megadiverse_countries}}'', Venezuela is listed as having a higher biodiversity than Colombia.
    \item \textbf{Q6:} \textit{Which single man-eating animal is responsible for the most deaths ever?} Mosquitoes being responsible for the most deaths is somewhat common knowledge; however, they transmit deadly diseases instead of being man-eaters. Regarding the \textit{single deadliest man-eater}, the Champawat Tigress (a tigress living in Nepal and Northern India in the late 19th century) is the correct answer. Chatbots usually get it right, but they are prone to overlooking the man-eater aspect and include mosquitoes in their answers.
\end{itemize}

\newpage
\section{Descriptive Statistics for Behavioral Variables}
\label{descriptive_statistiscs}

Table~\ref{tab:behavioral_desc} shows the percentage of zero values, means, standard deviations, min-max values, and percentiles for behavioral variables.  Means are close to zero, especially in InteractionLen and AI-URLsNum, due to a significant number of zero values. 

Almost 60\% of the participants had not interacted with the chatbot at all. This represents the aggregated number of interactions for questions and contains participants who might have interacted with the chatbot only once. If we check each question individually, the number of zero interactions varies between 75-80\%. For URLs visited, the distribution of zeros is not as extreme. We observe that only 28\% of the participants did not check any other sources at all. On the question level, the percentage varies between 36-38\%. 

The distribution of zeros for AI URLs follows a similar pattern to chatbot interactions. Aggregated, we see that 59\% of the participants did not visit any other AI sources. On the question-level, the number varies between 76-82\%. Non-AI URLs have less zero-inflation, as only 32\% of the participants did not visit any other Non-AI source. On the question-level, the number varies between 44-46\%.  Using different types of sources is not mutually exclusive: We see that only 13\% of the participants did not interact with the chatbot or did not visit any URLs at all. This number varies between 27-31\% on the question level.

\% Zero and maximum value for averages might look unorthodox since they are smaller than the question values. \% Zero values are lower because someone interacting with a chatbot or clicking on a link only once becomes non-zero when aggregated over questions. Maximum values are lower in aggregates because having a high value for a single question does not mean a participant would also have a higher value for all questions. On the contrary, too much engagement increases fatigue and lowers the likelihood of interacting at the same pace in the next question.

    \begin{table}[htbp]
        \centering
        \begin{adjustbox}{max width=\textwidth}
        \begin{tabular}{llrrrrrrrr}
        \toprule
        \textbf{Variable} & \textbf{Item} & \textbf{\% Zero} & \textbf{Mean} & \textbf{SD} & \textbf{Min} & \textbf{P25} & \textbf{Median} & \textbf{P75} & \textbf{Max} \\
        \midrule
        \multirow{7}{*}{InteractionLen}
         & Q1 & 77.4 & 0.42 & 1.02 & 0.0 & 0.0 & 0.0 & 0.0 & 9.0 \\
         & Q2 & 77.8 & 0.38 & 0.95 & 0.0 & 0.0 & 0.0 & 0.0 & 7.0 \\
         & Q3 & 76.4 & 0.32 & 0.71 & 0.0 & 0.0 & 0.0 & 0.0 & 5.0 \\
         & Q4 & 77.3 & 0.31 & 0.65 & 0.0 & 0.0 & 0.0 & 0.0 & 3.0 \\
         & Q5 & 79.3 & 0.28 & 0.67 & 0.0 & 0.0 & 0.0 & 0.0 & 5.0 \\
         & Q6 & 78.7 & 0.28 & 0.69 & 0.0 & 0.0 & 0.0 & 0.0 & 4.0 \\
         & Avg. & 59.7 & 0.33 & 0.59 & 0.0 & 0.0 & 0.0 & 0.5 & 3.5 \\
        \midrule
        \multirow{7}{*}{AI-URLsNum}
         & Q1 & 76.9 & 0.74 & 1.89 & 0.0 & 0.0 & 0.0 & 0.0 & 11.0 \\
         & Q2 & 76.4 & 0.59 & 1.69 & 0.0 & 0.0 & 0.0 & 0.0 & 14.0 \\
         & Q3 & 77.9 & 0.65 & 1.66 & 0.0 & 0.0 & 0.0 & 0.0 & 11.0 \\
         & Q4 & 78.4 & 0.72 & 2.55 & 0.0 & 0.0 & 0.0 & 0.0 & 28.0 \\
         & Q5 & 81.9 & 0.49 & 1.40 & 0.0 & 0.0 & 0.0 & 0.0 & 10.0 \\
         & Q6 & 75.4 & 0.55 & 1.54 & 0.0 & 0.0 & 0.0 & 0.0 & 14.0 \\
         & Avg. & 59.3 & 0.63 & 1.36 & 0.0 & 0.0 & 0.0 & 0.7 & 8.2 \\
        \midrule
        \multirow{7}{*}{NonAI-URLsNum}
         & Q1 & 44.2 & 1.73 & 2.50 & 0.0 & 0.0 & 1.0 & 2.0 & 17.0 \\
         & Q2 & 45.7 & 1.85 & 3.10 & 0.0 & 0.0 & 1.0 & 2.0 & 27.0 \\
         & Q3 & 44.2 & 1.67 & 2.34 & 0.0 & 0.0 & 1.0 & 2.0 & 13.0 \\
         & Q4 & 45.7 & 1.91 & 3.28 & 0.0 & 0.0 & 1.0 & 2.0 & 19.0 \\
         & Q5 & 44.7 & 1.56 & 2.66 & 0.0 & 0.0 & 1.0 & 2.0 & 17.0 \\
         & Q6 & 44.7 & 1.52 & 2.55 & 0.0 & 0.0 & 1.0 & 2.0 & 17.0 \\
         & Avg. & 31.7 & 1.71 & 2.09 & 0.0 & 0.0 & 1.2 & 2.5 & 11.8 \\
        \midrule
        \multirow{7}{*}{URLsNum}
         & Q1 & 37.7 & 2.47 & 3.51 & 0.0 & 0.0 & 2.0 & 3.0 & 24.0 \\
         & Q2 & 37.7 & 2.44 & 3.85 & 0.0 & 0.0 & 1.0 & 3.0 & 31.0 \\
         & Q3 & 36.2 & 2.32 & 3.22 & 0.0 & 0.0 & 1.0 & 3.0 & 21.0 \\
         & Q4 & 37.7 & 2.64 & 4.64 & 0.0 & 0.0 & 1.0 & 3.0 & 29.0 \\
         & Q5 & 38.2 & 2.05 & 3.14 & 0.0 & 0.0 & 1.0 & 3.0 & 17.0 \\
         & Q6 & 37.7 & 2.07 & 3.25 & 0.0 & 0.0 & 1.0 & 3.0 & 23.0 \\
         & Avg. & 27.6 & 2.33 & 2.76 & 0.0 & 0.0 & 1.7 & 3.2 & 14.3 \\
        \midrule
        \multirow{7}{*}{InteractionLen + URLsNum}
         & Q1 & 31.1 &  &  &  &  &  &  &  \\
         & Q2 & 27.6 &  &  &  &  &  &  &  \\
         & Q3 & 27.1 &  &  &  &  &  &  & \\
         & Q4 & 29.1 &  &  &  &  &  &  & \\
         & Q5 & 28.1 &  &  &  &  &  &  &  \\
         & Q6 & 27.1 &  &  &  &  & &  &  \\
         & Avg. & 13.0 &  &  &  &  &  &  &  \\
        \bottomrule
        \end{tabular}
        \end{adjustbox}
        \caption{Descriptive statistics for the number of turns in the chatbot interaction (InteractionLen) and number of URLs visited (URLsNum, AI-URLsNum, and NonAI-URLsNum). The averages represent the mean of all six questions. \% Zero represents the percentage of zero values. InteractionLen + URLsNum is included to report the percentage of participants who did not interact with any source at all.}
        \label{tab:behavioral_desc}
    \end{table}

\newpage
\section{Survey Items}

    \begin{table}[htbp]
        \centering
        \scriptsize
        \begin{tabular}{llcccc}
        \toprule
        \textbf{Item} & \textbf{Scale} & \textbf{M} & \textbf{SD} & \textbf{Cronbach's $\alpha$} & \textbf{KMO} \\
        \midrule
        \textbf{Perceived Warmth} & \makecell[l]{7-point Likert \\ Agreement} & 5.22 & 1.22 & 0.94 & 0.84 \\
        \midrule
        \makecell[l]{The answers the chatbot \\ gave were warm.} \\
        \makecell[l]{The answers the chatbot \\ gave were friendly.} \\
        \makecell[l]{The answers the chatbot \\ gave were kind.} \\
        \makecell[l]{The answers the chatbot \\ gave were enthusiastic.} \\
        \midrule
        \textbf{Trust} & \makecell[l]{7-point Likert \\ Agreement} & 3.43 & 1.15 & 0.93 & 0.85 \\
        \midrule
        AI chatbots are reliable. \\ 
        I can trust AI chatbots. \\
        AI chatbots have integrity. \\
        AI chatbots are dependable. \\
        \midrule
        \textbf{Chatbot Literacy} & \makecell[l]{5-point Familiarity: \\ 1: None, 5: Fully} & 2.66 & 0.91 & 0.85 & 0.81 \\
        \midrule
        Generative AI \\
        Token \\
        Hallucination \\
        Large Language Model \\
        Transformer \\
        Retrieval-Augmented Generation \\
        Prompt \\
        \bottomrule
        \end{tabular}
    \caption{Survey items.}
    \label{tab:survey_items}
    \end{table}

\newpage
\section{Question-Level Differences in Perception and Performance}
\label{question_statistics}

\begin{table}[htbp]
    \centering
    \begin{tabular}{l cc cc cc}
     & \multicolumn{6}{c}{} \\
    \cmidrule(lr){2-7}
    \textbf{Question} 
    & \multicolumn{2}{c}{\textbf{Difficulty\textsubscript{chatbot}}} 
    & \multicolumn{2}{c}{\textbf{Confidence}} 
    & \multicolumn{2}{c}{\textbf{Familiarity}} \\
    \cmidrule(lr){2-3} \cmidrule(lr){4-5} \cmidrule(lr){6-7}
     & Mean & SD & Mean & SD & Mean & SD \\
    \midrule
    Q1 & 4.02 & 1.96 & 6.98 & 2.68 & 1.73 & 1.25 \\
    Q2 & 3.40 & 1.92 & 7.59 & 2.39 & 1.94 & 1.40 \\
    Q3 & 3.08 & 1.87 & 7.47 & 2.68 & 2.10 & 1.52 \\
    Q4 & 3.43 & 1.93 & 7.59 & 2.49 & 2.22 & 1.61 \\
    Q5 & 3.05 & 1.91 & 8.02 & 2.16 & 2.20 & 1.49 \\
    Q6 & 3.30 & 2.00 & 7.96 & 2.26 & 2.28 & 1.66 \\
    \bottomrule
    \end{tabular}
    \caption{Participant perceptions towards questions.}
    \label{tab:question_stats}
\end{table}

\begin{figure}[htbp]
    \centering
    \includegraphics[width=\linewidth]{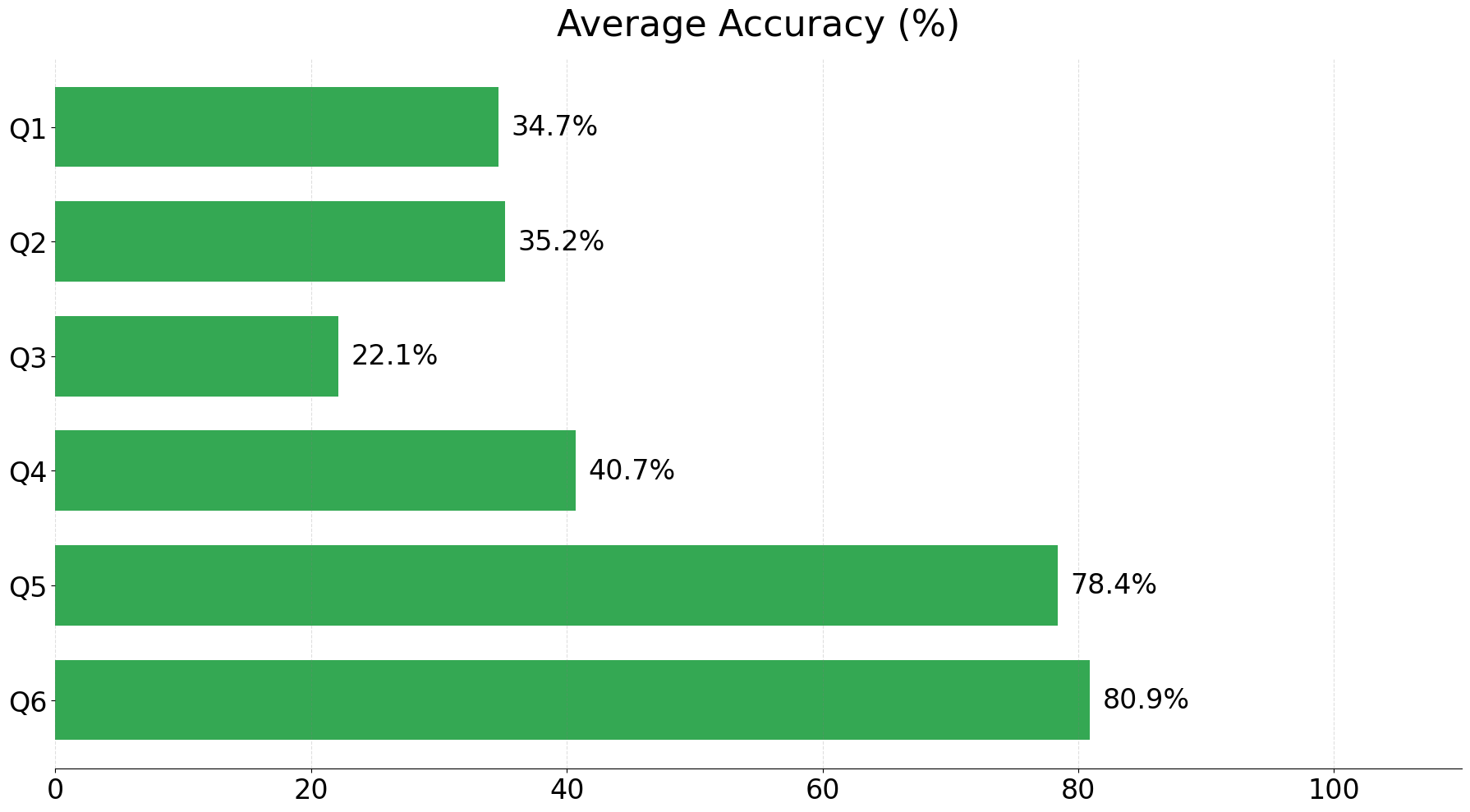}
    \caption{Average participant accuracy per question.}
    \label{fig:accuracy}
\end{figure}

Questions are perceived differently by the participants, as evident from Table~\ref{tab:question_stats}. That being said, perceptions don't align well with accuracy, at least not for all questions. Figure~\ref{fig:accuracy} shows that Q5 and Q6 were easy for the participants, while Q3 was the toughest question by far. Confidence for Q5 and Q6 was higher compared to other questions, aligning with participant accuracy. Confidence and familiarity for Q1 were the lowest, which also aligns with the low average accuracy. The remaining questions don't follow a similar pattern, specifically, Q3. We believe this is due to the challenging nature of Q3, as explained in \ref{question_pitfalls}.

\begin{figure}[htbp]
    \centering
    \includegraphics[width=\linewidth]{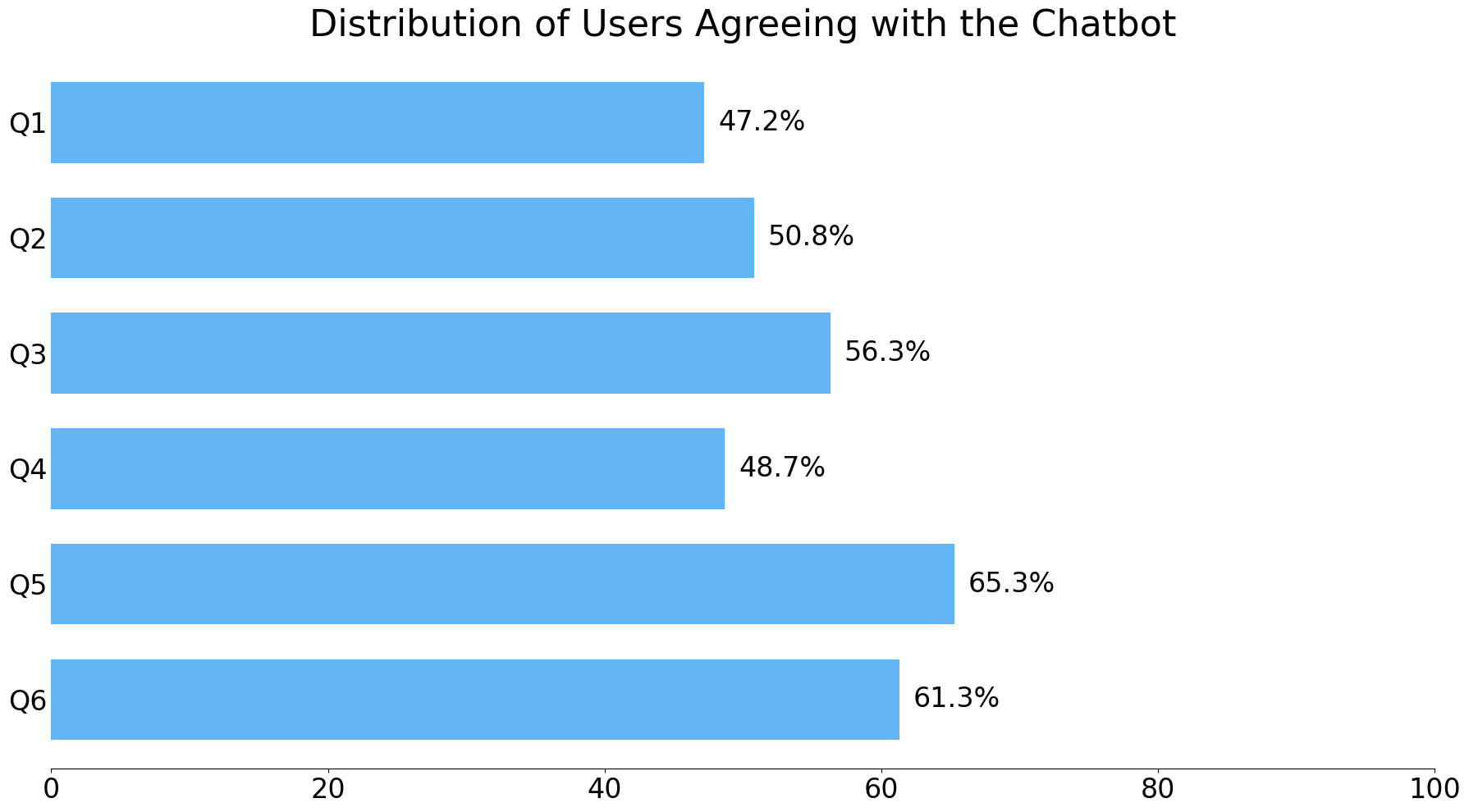}
    \caption{Percentage of participants agreeing with the chatbot per question.}
    \label{fig:agreement}
\end{figure}

Agreement levels between questions do not show drastic differences (see Figure~\ref{fig:agreement}). Question ID did not vary meaningfully as a random effect for the agreement model, and the plot validates our finding. This aligns with our hypothesis that behavior towards chatbots, such as agreement, is mostly participant-dependent, and the question characteristic does not play a meaningful role.

\begin{figure}[htbp]
    \centering
    \includegraphics[width=\linewidth]{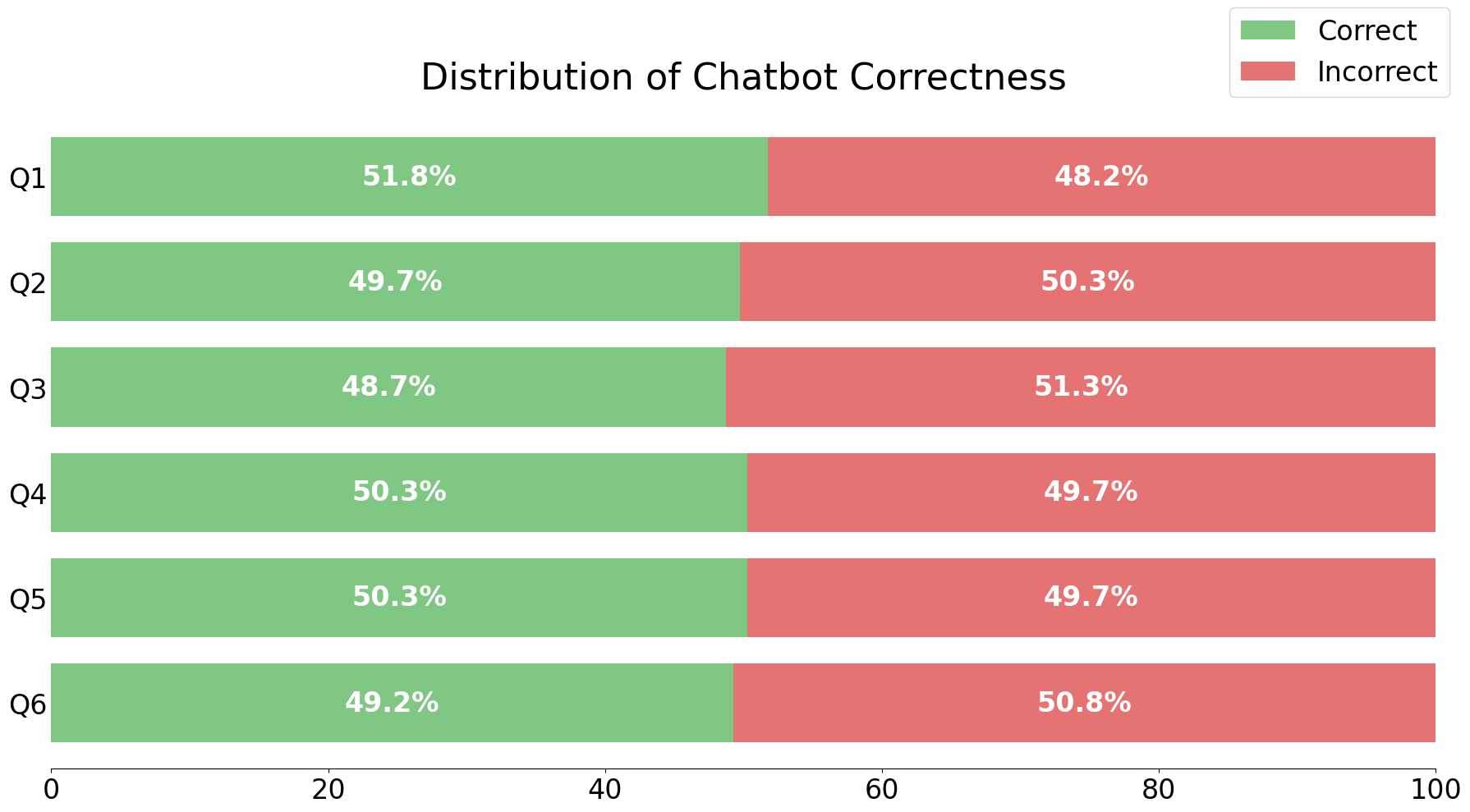}
    \caption{Correctness distribution of the chatbot's initial answer per question.}
    \label{fig:correctness}
\end{figure}

Initially, we aimed for the correctness of initial chatbot answers to be distributed equally, as 50\%-50.\% However, after filtering some participants, the distributions have shifted. Figure~\ref{fig:correctness} shows that for all questions, the correct/incorrect distribution is close to 50-50. At worst, for Q1, the distribution is tipped towards correct answers as 51.8\% correct to 48.2\% incorrect. Given the significant differences in accuracy and agreement levels between questions, we concluded that the marginal imbalances in the correctness distributions were not significant enough for an intervention. Questions were included as random-effects in both the accuracy and agreement models, alongside correctness as a control variable to account for any influence distribution imbalance might have.

\newpage
\section{Participant Demographics}
\label{participant_demographics}

\begin{table}[htbp]
    \centering

    \begin{tabular}{l r}
    \toprule
    Education Level & N \\
    \midrule
    Primary or lower secondary education & 1 \\
    High school & 41 \\
    Vocational or technical training & 26 \\
    BA/BS & 86 \\
    MA/MS & 37 \\
    PhD & 8 \\
    \bottomrule
    \end{tabular}

    \vspace{1em}

    \begin{tabular}{l r}
    \toprule
    Usage Frequency & N  \\
    \midrule
    3+ times a day & 46 \\
    1--2 times a day & 58 \\
    Weekly & 56 \\
    Monthly & 20 \\
    Less than once a month & 3 \\
    Tried a couple of times, do not use regularly & 15 \\
    Never & 1 \\
    \bottomrule
    \end{tabular}
    \caption{Distribution of education level and chatbot usage frequency of participants.}
    \label{tab:participant_characteristics}
\end{table}

\begin{figure}[htbp]
    \centering
    \includegraphics[width=0.6\linewidth]{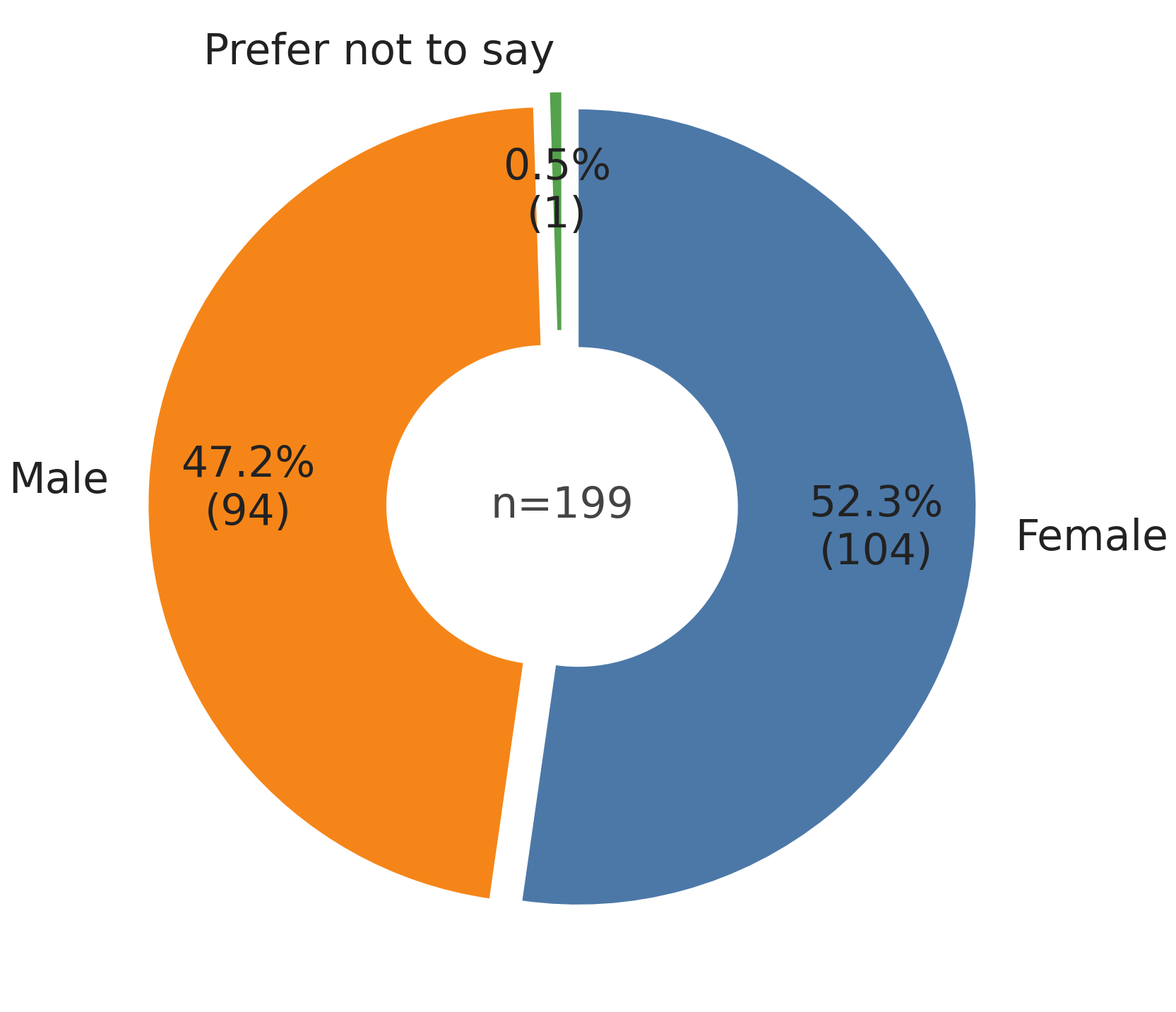}
    \caption{Gender distribution of participants.}
    \label{fig:gender_distribution}
\end{figure}

\begin{figure}[htbp]
    \centering
    \includegraphics[width=0.8\linewidth]{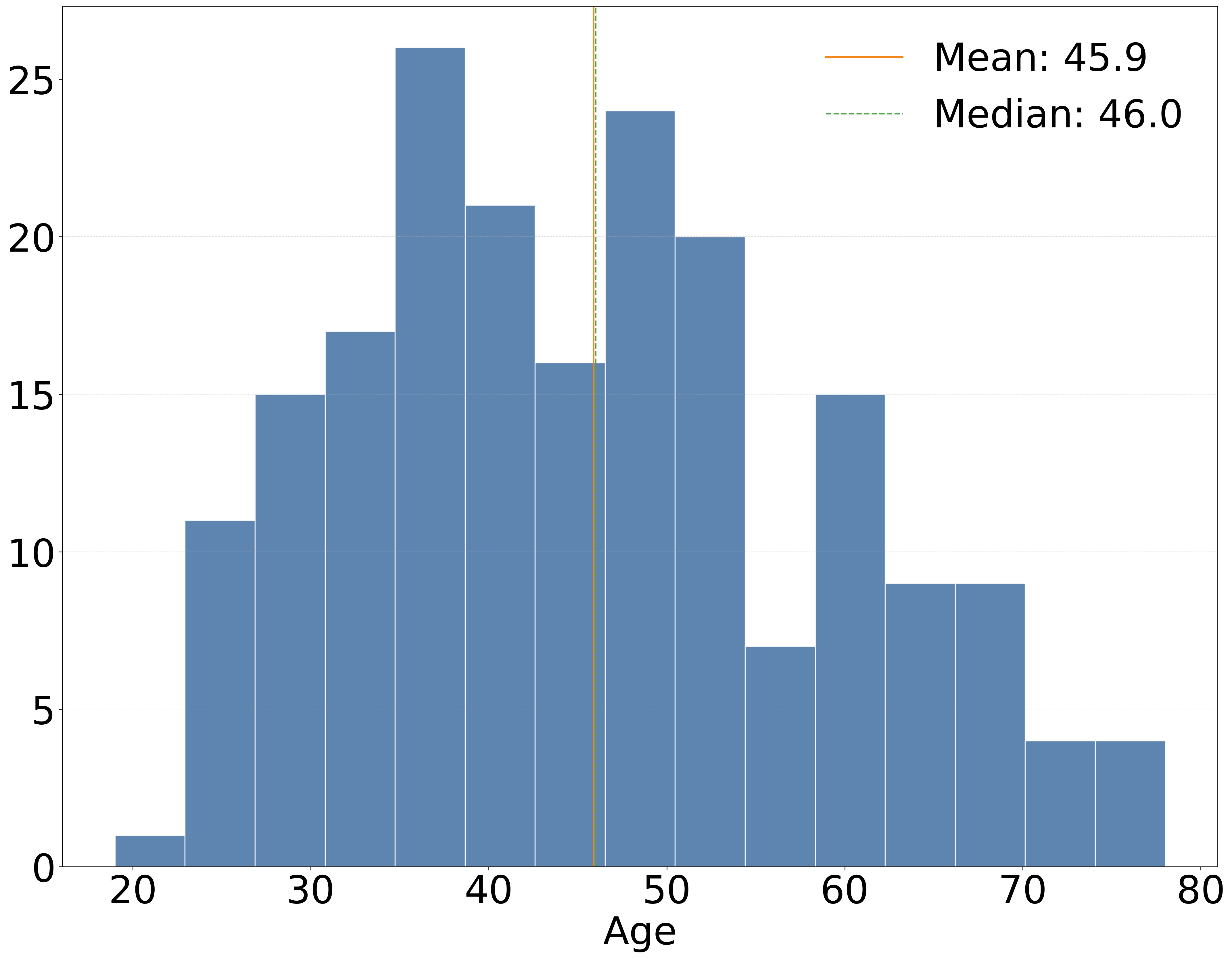}
    \caption{Age distribution of participants.}
    \label{fig:age_distribution}
\end{figure}

\newpage

\bibliographystyle{elsarticle-harv} 
\bibliography{bibliography}

\end{document}